\documentclass[letter]{aa} % for the letters 
\usepackage{graphicx}
\usepackage{txfonts}
\usepackage[pdfpagelabels=false]{hyperref}
\hypersetup{colorlinks=true,linkcolor=red,citecolor=blue,filecolor=blue,urlcolor=blue}

\DeclareRobustCommand{\VAN}[3]{#2}
\let\VANthebibliography\thebibliography
\def\thebibliography{\DeclareRobustCommand{\VAN}[3]{##3}\VANthebibliography}
\numberwithin{equation}{section}    %Number equation by section

\usepackage{booktabs}   %for a type of table
\usepackage{mhchem} %For the chemical elements
\usepackage{comment} 
\numberwithin{equation}{section}    %Number equation by section
\usepackage{caption}
\usepackage{subcaption}
\usepackage[referable,para]{threeparttablex}
\usepackage{float} %To use H in the figures' option
\usepackage{placeins}% [section] In order to use \FloatBarrier
\newcommand{\kepler}{\textit{Kepler}\xspace}

\newcommand{\numax}{\ensuremath{\nu_{\mathrm{max}}}\xspace}
\newcommand{\msol}{\ensuremath{\mathrm{M}_\odot}\xspace}

\newcommand{\teff}{\ensuremath{{{T}_{\mathrm{eff}}}}\xspace}

\begin{document} 

\title{Red Horizontal Branch stars: an asteroseismic perspective}

\author{Massimiliano Matteuzzi\inst{\ref{inst1},\ref{inst2}}\thanks{E-mail: \href{mailto:massimilia.matteuzz2@unibo.it}{massimilia.matteuzz2@unibo.it}} \and
Josefina Montalb\'an\inst{\ref{inst1},\ref{inst3}}
\and
Andrea Miglio\inst{\ref{inst1},\ref{inst2},\ref{inst3}}
\and
Mathieu Vrard\inst{\ref{inst4}}
\and
Giada Casali\inst{\ref{inst1},\ref{inst2}}
\and \\
Amalie Stokholm\inst{\ref{inst1},\ref{inst5}}
\and
Marco Tailo\inst{\ref{inst1}}
\and
Warrick Ball\inst{\ref{inst3}}
\and
Walter E. van Rossem\inst{\ref{inst3},\ref{inst5}}
\and
Marica Valentini\inst{\ref{inst6}}
}

\institute{Department of Physics \& Astronomy "Augusto Righi", University of Bologna, via Gobetti 93/2, 40129 Bologna, Italy\label{inst1}
\and
INAF-Astrophysics and Space Science Observatory of Bologna, via Gobetti 93/3, 40129 Bologna, Italy\label{inst2}
\and
School of Physics and Astronomy, University of Birmingham, Edgbaston, Birmingham B15 2TT, UK\label{inst3}
\and
Department of Astronomy, The Ohio State University, Columbus, OH 43210, USA\label{inst4}
\and
Stellar Astrophysics Centre, Department of Physics and Astronomy, Aarhus University, Ny Munkegade 120, DK-8000 Aarhus C, Denmark\label{inst5}
\and
Leibniz-Institut für Astrophysik Potsdam, An der Sternwarte 16, Potsdam, 14482, Germany\label{inst6}
}

\abstract{
%Precise masses of red giant stars enable a robust inference of their ages. There are, however, core-helium-burning (CHeB) stars with such an inaccurate age estimation that they seem significantly older than the Universe. They provide candidates stripped stars or stars that have lost more mass than expected. In our study, we focus on a group of solar-like oscillators in the {\it Kepler} field that are located between the red edge of the RR Lyrae and the red clump (RC) in the HR diagram. We expect them to have lost more mass than a RC star, but less than a subdwarf B (sdB) star. Their pulsation patterns are more complex than RC stars, with a higher number of visible mixed modes. We generate realistic artificial power spectral densities simulating the $\approx 4$ yr baseline of the {\it Kepler} mission. We confirm that these stars are characterised by a rather \emph{extreme} coupling (i.e. $q>0.4$), leading to a large number of detectable mixed modes. These stars are therefore consistent with simulated CHeB stars that contain $\approx 0.5$ \msol of Helium in the core and a small envelope left ($\approx 0.1-0.2$ \msol). This holds detailed information about the structural properties of stars that probably underwent mass stripping processes, hence it can potentially provide another piece of the puzzle in the sequence between RC and sdB stars or other stripped stars.

Robust age estimates of red giant stars are now possible thanks to the precise inference of their mass based on asteroseismic constraints. However, there are cases where such age estimates can be highly precise yet very inaccurate. An example is giants that have undergone mass loss or mass transfer events that have significantly altered their mass. In this context, stars with “apparent” ages significantly higher than the age of the Universe are candidates as stripped stars, or stars that have lost more mass than expected, most likely via interaction with a companion star, or because of the poorly understood mass-loss mechanism along the red-giant branch. 

In this work we identify examples of such objects among red giants observed by \kepler, both at low ([Fe/H] $\lesssim -0.5 $) and solar metallicity. By modelling their structure and pulsation spectra, we find a consistent picture confirming that these are indeed low-mass objects consisting of a He core of $\approx 0.5$~\msol and an envelope of $\approx 0.1-0.2$~\msol. Moreover, we find that these stars are characterised by a rather extreme coupling ($q \gtrsim 0.4$) between the pressure-mode and gravity-mode cavities, i.e. much higher than the typical value for red clump stars, providing thus a direct seismic signature of their peculiar structure. 

The complex pulsation spectra of these objects, if observed with sufficient frequency resolution, hold detailed information about the structural properties of likely products of mass stripping, hence can potentially shed light on their formation mechanism. 
On the other hand, our tests highlight the difficulties associated with measuring reliably the large frequency separation, especially  in shorter datasets, with impact on the reliability of the inferred masses and ages of low-mass Red Clump stars with e.g. K2 or TESS data. 
}


\titlerunning{rHBs as viewed by asteroseismology}
\authorrunning{Matteuzzi et al.}
\maketitle
%%%%%%%%%%%%%%%%%%%%%%%%%%%%%%%%%%%%%%%%%%%%%%%%%%%%%%%%%%%%%%%%%%%%%%%%%%%%%%%%%%%%%%%%%%
\section{Introduction}
\label{sec:intro}
It is widely accepted that the large range in color shown by low mass stars in the central He-burning phase, called Horizontal Branch (HB), is mainly due to variations in the efficiency of the H-burning shell, hence, to the mass of the H-envelope remaining around a He-core of $\simeq 0.5$ \msol \citep[e.g.][]{SalarisCassisi2006}. In a colour-magnitude diagram (CMD), low-mass core-He-burning (CHeB) stars appear distributed in both bluer and redder colours than the RR Lyrae instability strip ($\rm RRL-IS$). Those located between the RR Lyrae and the red-clump \citep[RC; e.g.][]{2016ARA&A..54...95G} are called rHB (red Horizontal Branch) stars, and they have a H-rich envelope of $\approx 0.1-0.2$ \msol \citep[e.g.][]{1989upsf.conf..103R,2008A&A...487..185V,2016ARA&A..54...95G,2020MNRAS.498.5745T}. This HB component has been clearly observed in globular clusters of different metallicity and age \citep[e.g.][]{1988AJ.....96..588A,1989AJ.....97.1360S,2009Ap&SS.320..261C, 2020MNRAS.498.5745T}, however, rHB objects also exist in the field. While their identification is challenging, their census has been considered extremely important for tracing old stellar populations in the Milky Way \citep[MW; e.g.][]{Kaempf2005,Chen2010,Chen2011}. Although mainly associated with stars of low/intermediate metallicity, corresponding to the thick disc and halo population, spectroscopic studies \citep[e.g.][]{Afsar2012A,Afsar2018} have shown that rHB stars are also present in the metal-rich component of the MW. This suggests that the progenitors of these objects have followed a non-standard evolution with significant mass loss, or envelope stripping due to binary interactions. Signs of significant mass loss have been revealed in red giants observed by the \kepler space telescope \citep{Borucki2010}, in the field and in the open cluster NGC~6819 \citep[e.g.][]{2017MNRAS.472..979H,2021MNRAS.507..496B,2022NatAs...6..673L}.

\begin{figure*}
%%%%%%%%%%%%%%%%%%%%%%%%%%%%%%%%%%%%%%%%%%%%%%%%%%%%%%%%%%
\centering
\includegraphics[width=\columnwidth]{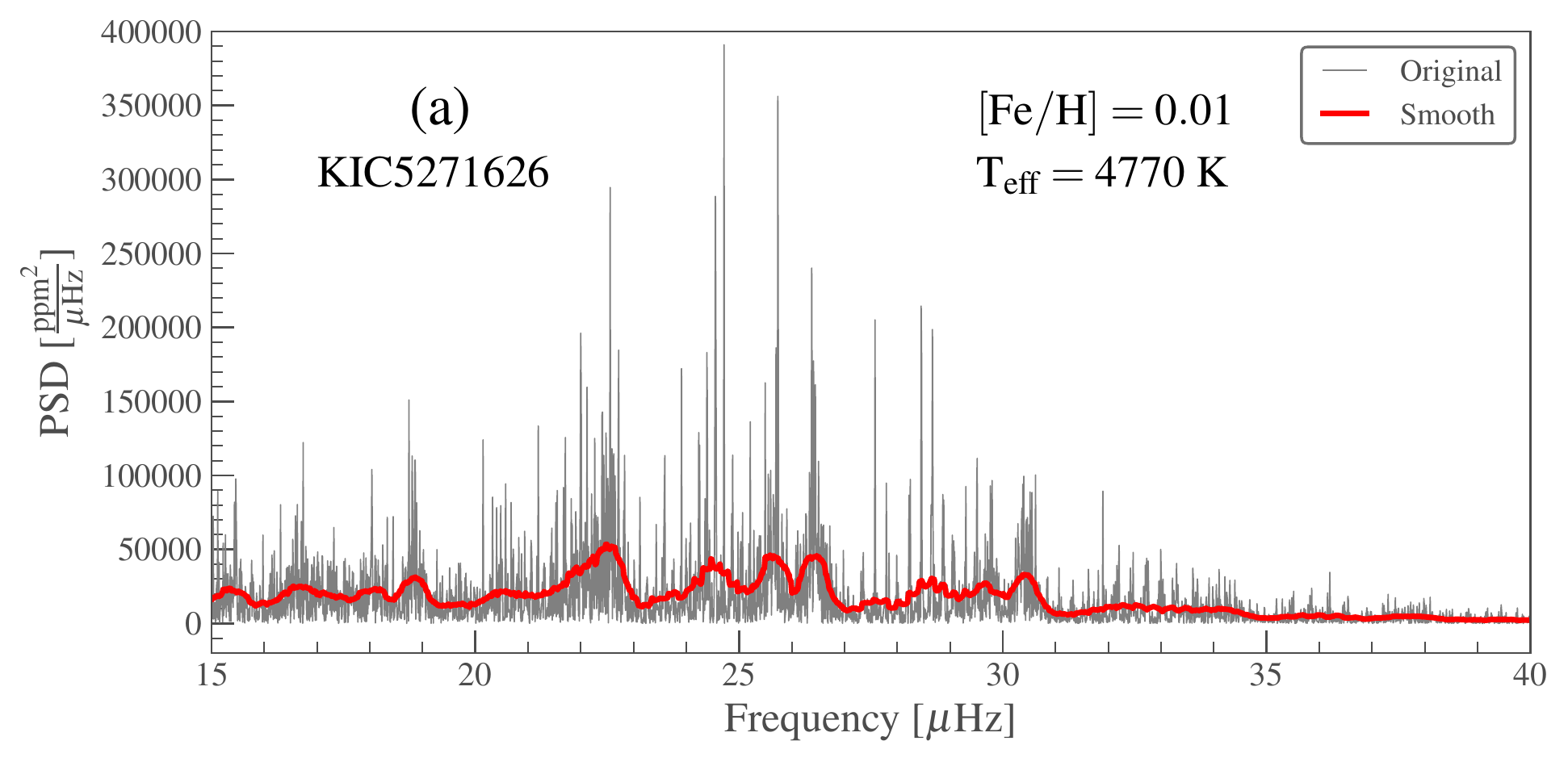}%width=\columnwidth
\includegraphics[width=\columnwidth]{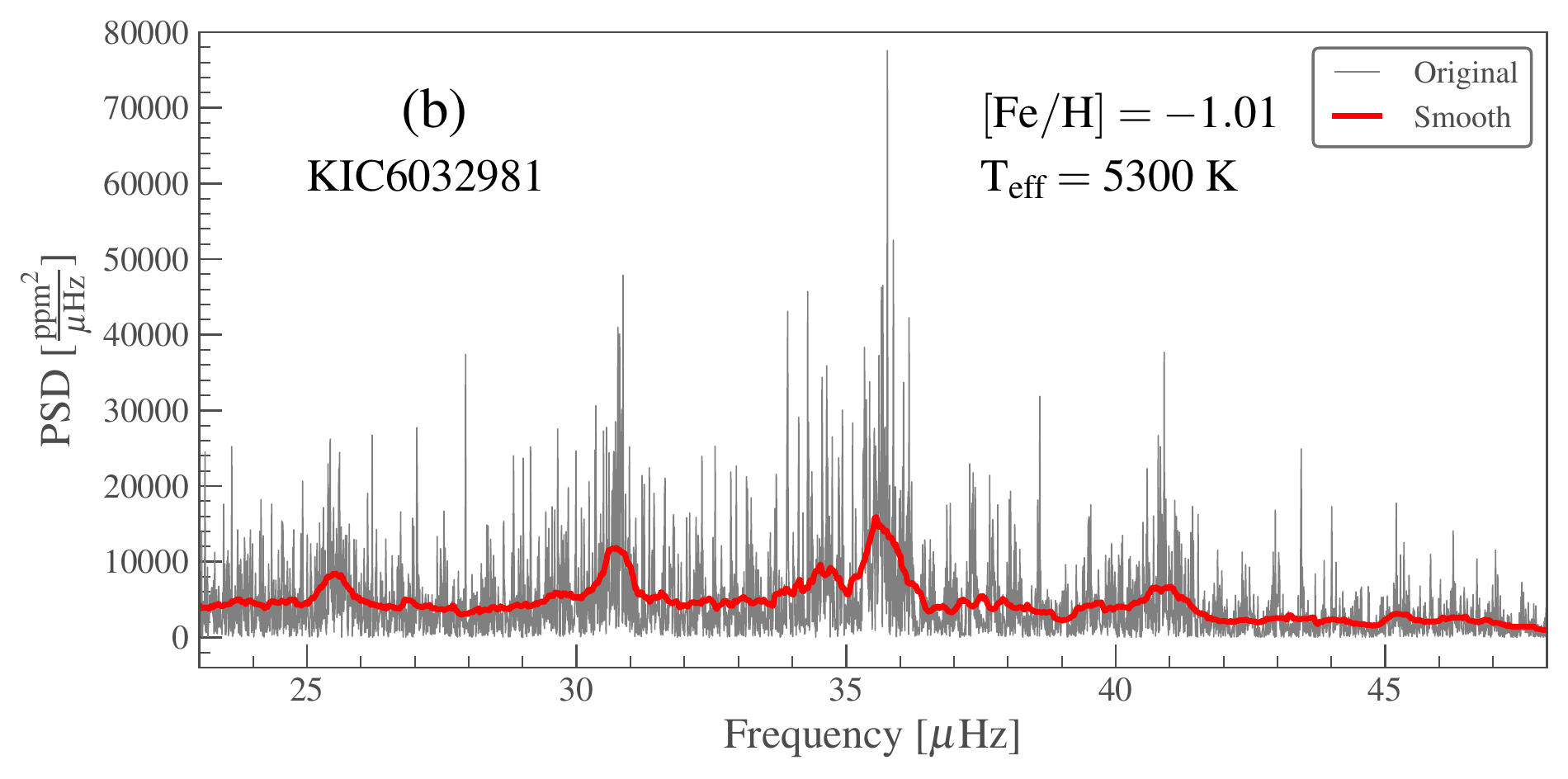}
\includegraphics[width=\columnwidth]{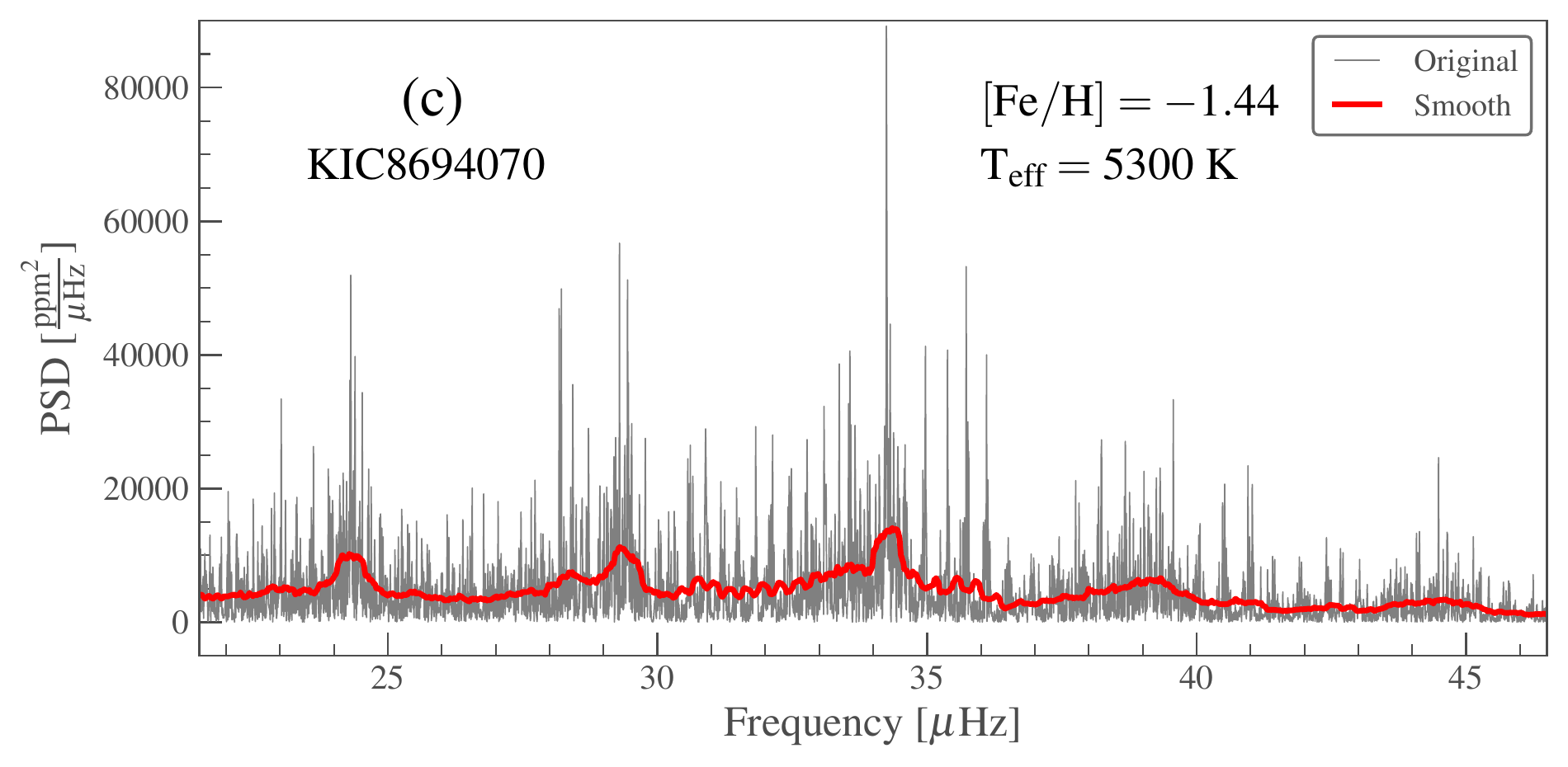}
\includegraphics[width=\columnwidth]{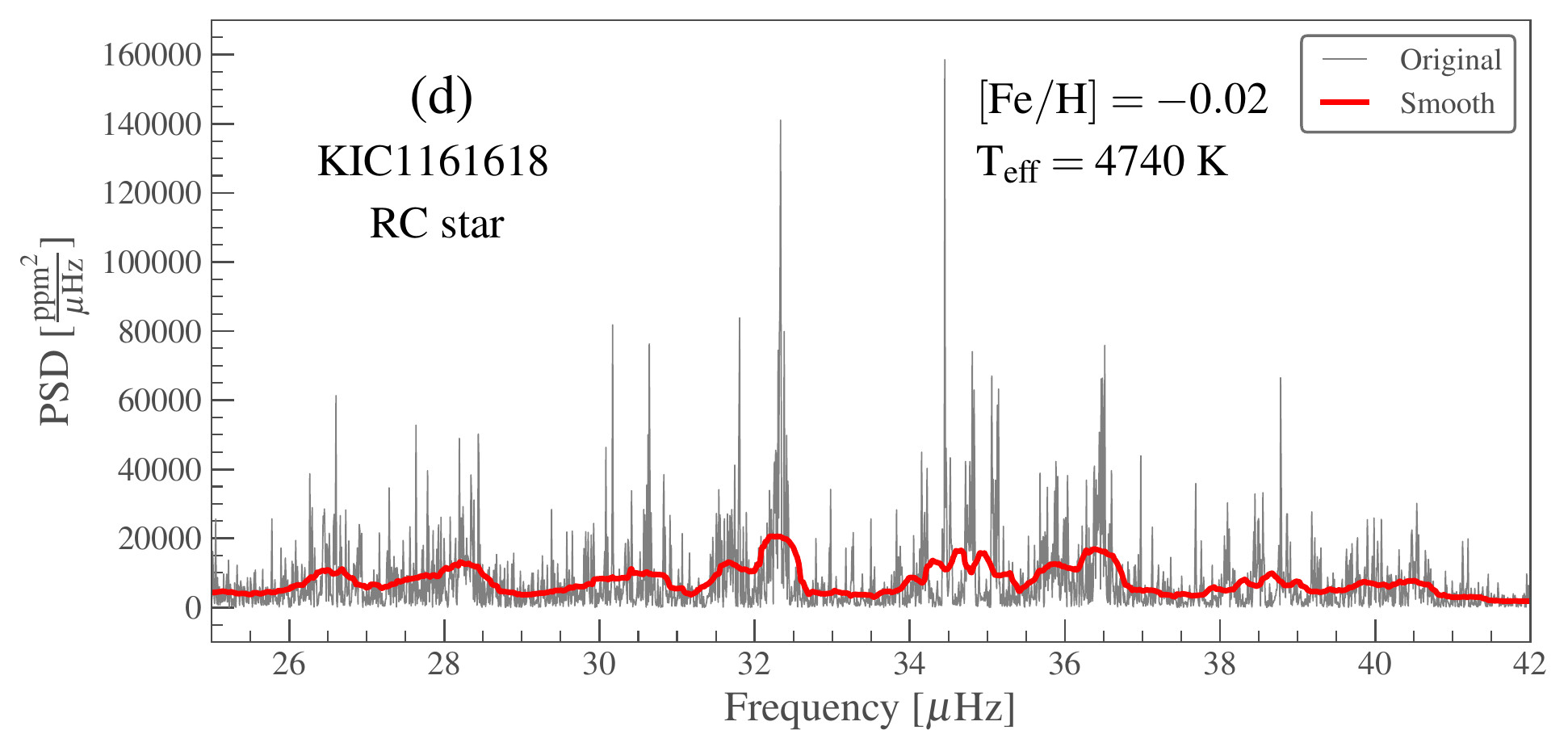}%RC star
\includegraphics[width=\columnwidth]{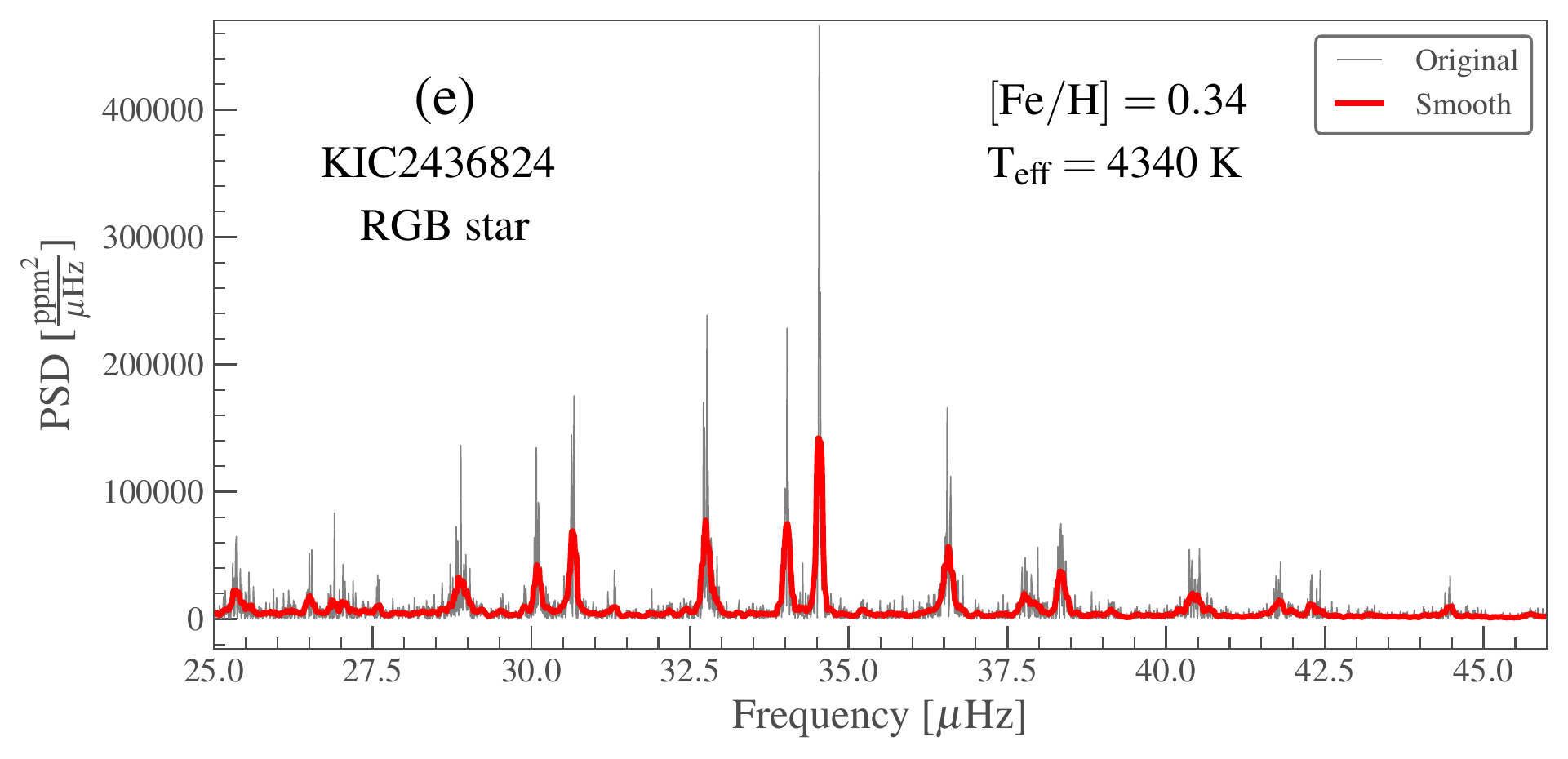}%RGB star
\caption{PSD for five low-mass red giants (grey lines in the five panels) observed by \kepler. Panels (a), (b), and (c) show the three low-mass CHeB stars KIC5271626, KIC6032981, and KIC8694070 (first three rows in Table~\ref{tab:sample} and colored stars in Fig.~\ref{fig:HR}). Panels (d) and (e) show the RC star KIC1161618 and the RGB star KIC2436824, for comparison. All the five panels contain a smoothed PSD (red lines) computed with a box kernel of width 0.5 $\mu$Hz in panels (a), (b), (c), (d), and of width 0.1 $\mu$Hz in panel (e).}
\label{fig:KICs}
\end{figure*}

Stellar evolution models predict different structures for rHB and RC stars, with the latter having a similar He core, but a larger H envelope. We thus expect their seismic properties to be different. The exquisite precision achieved after $4$ years of {\it Kepler} observations has revealed oscillation spectra of red giants with an increasing level of complexity \citep[see][for a review]{ChaplinMiglio2013}: frequency patterns in red giant branch (RGB) stars similar to those found in main sequence stars \citep[Universal pattern,][]{2011A&A...525L...9M}, spectra of RC stars with "forests" of dipole modes around the nominal acoustic mode, but still with an evident regularity \citep[i.e.][]{2011Sci...332..205B}, and also "outlier" spectra with a larger number of visible modes over the whole frequency domain, which are hypothesised in this letter to belong to rHB stars (see Figure \ref{fig:KICs}). 

In this work we identify a small sample of 11 rHB candidates among the red giants in the \kepler field. Their global seismic properties and atmospheric parameters suggest that they are low-mass CHeB stars with low/intermediate and solar metallicity. Combining numerical simulations of stellar structure and evolution and stellar oscillations, we study the consistency between the location of our rHB candidates in the Hertzsprung–Russell diagram (HRD), their theoretically predicted internal structure and their oscillation spectra. Our rHB sample is presented in Section~\ref{sec:obs} and the theoretical models in Section~\ref{sec:anal}. Section~\ref{sec:results} discusses the properties of theoretical oscillation spectra of typical rHB and RC stars, as well as the comparison with observations. In Section~\ref{sec:concl} we summarise our findings. 

%%%%%%%%%%%%%%%%%%%%%%%%%%%%%%%%%%%%%%%%%%%%%%%%%%%%%%%%%%
%%%%%%%%%%%%%%%%%%%%%%%%%%%%%%%%%%%%%%%%%%%%%%%%%%%%%%%%%%%%%%%%%%%%%%%%%%%%%%%%%%%%%%

%%%%%%%%%%%%%%%%%%%%%%%%%%%%%%%%%%%%%%%%%%%%%%%%%%%%%%%%%%%%%%%%%%%%%%%%%%%%%%%%%%%%%%
\section{Observational data}
\label{sec:obs}
In addition to KIC4937011, a 0.71 \msol CHeB star belonging to the open cluster NGC~6819 \citep[see][]{2017MNRAS.472..979H} which has a turn-off mass of $\sim 1.6$ \msol, we found 11 red giants in the \kepler database\footnote{\href{https://archive.stsci.edu/missions-and-data/kepler}{https://archive.stsci.edu/missions-and-data/kepler}} with peculiar power spectral density (PSD). While their global seismic parameters (mean large frequency separation $\left<\Delta \nu\right>$, frequency of maximum power $\nu_{\rm max}$, and asymptotic period spacing of the dipole modes $\Delta \Pi_1$) are compatible with low-mass CHeB stars, they have complex oscillation spectra. They have, for instance, an unusually high number of observable dipole g/p mixed modes without the amplitude modulation around the p-like modes that is typically found in low-RGB and RC stars. This fact suggests that all dipole modes have a significant amplitude also in the outer region of the star, and hence, that g and p resonant cavities in these objects are strongly coupled.

The ability to transfer the energy of the mode from one cavity to the other, instead of remaining trapped mainly in one of them, is quantified by the coupling factor $q$ \citep[e.g.][]{1979PASJ...31...87S,2016PASJ...68..109T}. The analysis of \kepler light curves provides the seismic parameters mentioned above, as well as the value of the coupling factor $q$ \citep[e.g.][]{2016A&A...588A..87V,2017A&A...600A...1M, Mosser2018}. Theoretically, the parameter $q$ ranges from 0 (uncoupled) to 1 (completely coupled). All the stars in our sample have $q\gtrsim0.4$, while the median for RC stars is $\simeq 0.25-0.3$ \citep{2016A&A...588A..87V,2017A&A...600A...1M}.

On the other hand, the values of radial mode-linewidths ($\Gamma_0>0.2 \ \mu$Hz) are larger than the third quantile of the full sample of CHeB \textit{Kepler} stars \citep[median value $\Gamma_0=0.15\ \mu$Hz,][]{Vrard2018}. Both a high $q$ and a high $\Gamma_0$ contribute to increase the complexity of the spectra. Moreover, given the dependence of $\Gamma_0$ on the effective temperature \teff \citep[e.g.][]{Chaplin2009}, the quadrupole modes are more difficult to detect in the hotter metal-poor subsample than in the cooler metal-rich objects.

The seismic properties (\numax, $\langle\Delta\nu\rangle$, $\Delta\Pi_1$ and $q$) for our sample are reported in Table~\ref{tab:sample} and \ref{tab:sample_full}, together with the atmospheric parameters (\teff and chemical composition) from APOGEE-DR16/DR17 \citep[] []{2020ApJS..249....3A,2022ApJS..259...35A}. Around 25\% (3 out of 12) of the sample are metal-rich ($0 \le \rm [Fe/H]<0.3$) cool ($4600 \le T_{\rm eff}/\textit{K} \le 4800 $) stars, and the rest are low/intermediate metallicity ($-1.4<\rm [Fe/H]<-0.5$) stars with $5200 \le T_{\rm eff}/\textit{K} \le 5600 $, that is, belonging to the "classical" rHB.

Tables~\ref{tab:sample} and \ref{tab:sample_full} contain also the stellar luminosity derived using \textit{Gaia}-DR3 astrometry data (see Appendix \ref{app:a1} for details), and  an estimate of their mass. The latter can be derived from scaling relations involving atmospheric and global seismic parameters \citep[see e.g.][]{2012MNRAS.419.2077M}. Here we use the one combining $L$, \teff and \numax:

%%%%%%%%%%%%%%%%%%%%%%%%%%%%%%%%%%%%%%%%%%%%%%%%%%%%%%%%%%%%%%%%%%%%%%%%%%%%%%%%%%%%%%%%%%%%%%%%
\begin{equation}
\label{eq:mass_scaling}
\centering
    \frac{M}{M_\odot} = \left(\frac{T_{\rm eff,\odot}}{T_{\rm eff}}\right)^{3.5}\ \left(\frac{\nu_{\rm max}}{\nu_{\rm max,\odot}}\right) \ \left(\frac{L}{L_\odot}\right),
\end{equation}
%%%%%%%%%%%%%%%%%%%%%%%%%%%%%%%%%%%%%%%%%%%%%%%%%%%%%%%%%%%%%%%%%%%%%%%%%%%%%%%%%%%%%%%%%%%%%%%%

\noindent where the solar reference values are $T_{\rm eff,\odot} = 5777$ K, $\nu_{\rm max,\odot} = 3090 \ \mu$Hz \citep[][]{2011ApJ...743..143H}. The mass uncertainties are calculated in quadrature by considering an uncertainty of at least 50~K in \teff as estimated from an independent analysis of APOGEE spectra (see Appendix~\ref{app:a1}). In Appendix~\ref{app:a2} we also discuss the stellar mass values from a model-based corrected scaling relation involving \teff, $\langle\Delta\nu\rangle$ and \numax (Eq.~\ref{eq:mass_scaling_appendix}).

We notice that the mass of KIC~4937011 in Table~\ref{tab:sample_full} is that of \citet[][]{2017MNRAS.472..979H}, and its value is nevertheless compatible with our results obtained with Eq.~\ref{eq:mass_scaling} or \ref{eq:mass_scaling_appendix}.
All the objects in our sample are then very low-mass stars ($M \lesssim 0.8$~\msol) with a high coupling\footnote{We notice that stars in the CHeB stage could have multiple cavities in the inner part due to semi-convection. This could lead to bias when estimating $q$ from the fit of observations with the asymptotic relation for dipole modes \citep[e.g.][]{2022A&A...661A.139P}, thus it must be considered in future.} between p-mode and g-mode cavities. 

\begin{figure}
%%%%%%%%%%%%%%%%%%%%%%%%%%%%%%%%%%%%%%%%%%%%%%%%%%%%%%%%%%%%%%%%%%%%%%%%%%%%%%%%%%%%%%%%%%%%%%%%
\includegraphics[width=\columnwidth]{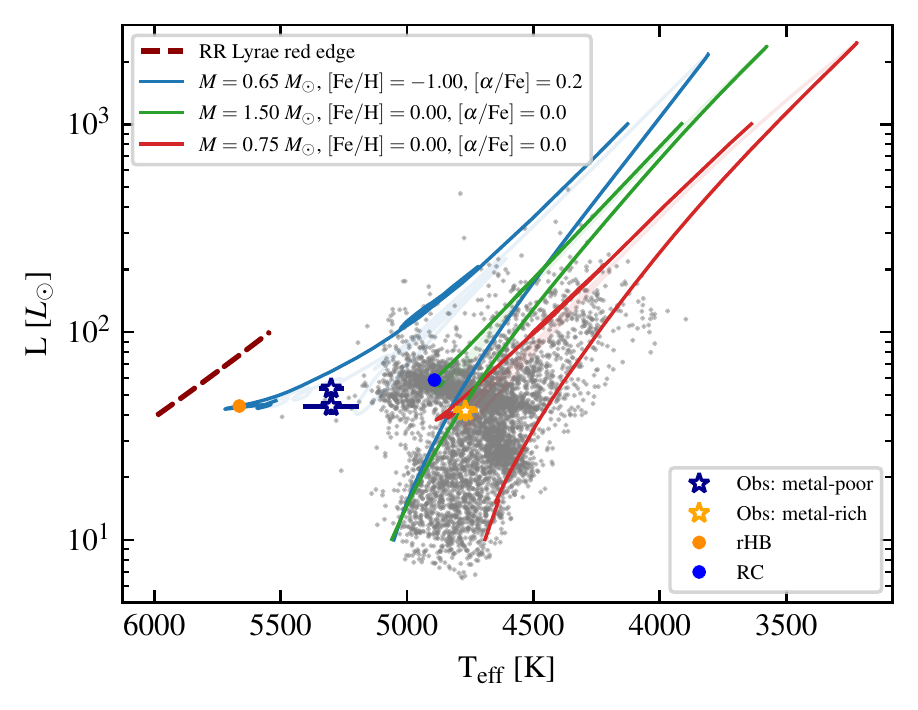}
%%%%%%%%%%%%%%%%%%%%%%%%%%%%%%%%%%%%%%%%%%%%%%%%%%%%%%%%%%%%%%%%%%%%%%%%%%%%%%%%%%%%%%%%%%%%%%%%
\caption{HRD of a sample of red giants in the \kepler field. The colored star symbols highlight the location of the first three rHB candidates in Table~\ref{tab:sample} and the grey dots in the background correspond to the \kepler-APOGEE sample in \citet[][]{2021A&A...645A..85M}. The blue and red lines represent the theoretical red giant evolutionary tracks (from the RGB phase until the first thermal pulse) of low-mass stars with two different chemical composition: $M=0.65 \ M_\odot$, $\rm [\alpha/Fe] = 0.2$, $\rm [Fe/H] = -1.00$ (blue), and for $M=0.75 \ M_\odot$, $\rm [\alpha/Fe] = 0$, $\rm [Fe/H] = 0$ (red). The green line is the evolutionary track for a 1.5 \msol with solar composition and the dashed red one is the red edge of the RRL-IS for the composition of the blue track \citep[see][]{2015ApJ...808...50M}. Solid orange and blue circles corresponds to our rHB and RC reference models, with a central He mass fraction $Y_{\rm c} \simeq 0.27$.}
\label{fig:HR}
\end{figure}

We select three stars (those in Table~\ref{tab:sample}) as representative of low-mass CHeB stars in different metallicity domains. Figure~\ref{fig:HR} shows these stars in an HRD, together with the \kepler-APOGEE red giant sample \citep[][grey dots]{2021A&A...645A..85M} and the red edge of the RRL-IS \citep[][dashed red line]{2015ApJ...808...50M}. The two metal-poor stars (blue star symbols) are located between the RRL-IS and the RC, as expected for rHB stars, while the metal-rich CHeB star (orange star symbol) appears in the region of the "ensemble" {\it Kepler}-RC. Its location is nevertheless redder than the RC at solar-metallicity, and hence it is well a rHB metal-rich star as  suggested also by its mass \citep[see also][]{2017MNRAS.472..979H} and oscillation spectra. 
As mentioned above, rHBs, especially those metal-rich, must have followed non-standard evolution to reach their current state within the age of the Universe. They are probably the progeny of strongly interacting binary systems. It has not been possible to confirm that hypothesis using the currently available \textit{Gaia}-DR3 astrometry data \citep[see][for the non-single star processing\footnote{We also checked the non-single star hypothesis using the \texttt{fidelity\_v2} table.}]{2022arXiv220605726H}, but we cannot exclude that they were part of binary systems in the past.

%%%%%%%%%%%%%%%%%%%%%%%%%%%%%%%%%%%%%%%%%%%%%%%%%%%%%%%%%%%%%%%%%%%%%%%%%%%%%%%%%%%%%%%%%%%%%%%%
%%%%%%%%%%%%%%%%%%%%%%%%%%%%%%%%%%%%%%%%%%%%%%%%%%%%%%%%%%%%%%%%%%%%%%%%%%%%%%%%%%%%%%%%%%%%%%%%

%%%%%%%%%%%%%%%%%%%%%%%%%%%%%%%%%%%%%%%%%%%%%%%%%%%%%%%%%%%%%%%%%%%%%%%%%%%%%%%%%%%%%%%%%%%%%%%%
\begin{table*}
\centering
\resizebox{\textwidth}{!}{%
%%%%%%%%%%%%%%%%%%%%%%%%%%%%%%%%%%%%%%%%%%%%%%%%%%%%%%%%%%%%%%%%%%%%%%%%%%%%%%%%%%%%%%%%%%%%%%%%
\begin{threeparttable}
\centering
%%%%%%%%%%%%%%%%%%%%%%%%%%%%%%%%%%%%%%%%%%%%%%%%%%%%%%%%%%%%%%%%%%%%%%%%%%%%%%%%%%%%%%%%%%%%%%%%
%\caption{Summary of the seismic and atmospheric properties for three rHB candidates of our sample (Sect.~\ref{sec:obs}). For each \kepler ID (KIC) we report the effective temperature $T_{\rm eff}$, $\rm [Fe/H]$, $\rm [\alpha/Fe]$ from APOGEE-DR17, or APOGEE-DR16 (one star tagged with a + in apex); mean large frequency separation $\left<\Delta \nu\right>$, and frequency of maximum power $\nu_{\rm max}$ calculated by us using the code in \citet[][]{2016AN....337..774D}, or \citet[][]{2018ApJS..236...42Y} data (one star tagged with a * in apex). The coupling factor $q$ and asymptotic period spacing of the dipole modes $\Delta \Pi_1$ as calculated from the stretched-period method \citep[see e.g.][]{2016A&A...588A..87V}. The current stellar mass $M$ is computed from Eq.~\ref{eq:mass_scaling}. The last two rows show the properties of a simulated rHB and RC star (Sect.~\ref{sec:anal}).}
\caption{Summary of the seismic and atmospheric properties for three rHB candidates of our sample (Sect.~\ref{sec:obs}).}
\label{tab:sample}
%%%%%%%%%%%%%%%%%%%%%%%%%%%%%%%%%%%%%%%%%%%%%%%%%%%%%%%%%%%%%%%%%%%%%%%%%%%%%%%%%%%%%%%%%%%%%%%%
\begin{tabular}{@{}lcllllllll@{}}
\toprule
KIC & $L$ [$\rm{L}_\odot$] & $T_{\rm eff}$ [K]  &  $\rm [Fe/H]$  & $\rm [\alpha/Fe]$  &  $\left<\Delta \nu\right>$ [$\mu$Hz]& $\nu_{\rm max}$ [$\mu$Hz]& $q$ & $\Delta \Pi_1$ [s] & $M$ [$\rm{M}_\odot$]\\ \midrule
$5271626^*$  & $42 \pm 4 $ & $4769 \pm 9$ & 0.03 &  0.01 & $3.91 \pm 0.05$& $25.1 \pm 0.5$ & 0.61 & $291.4 \pm 1.7$ & $0.66 \pm 0.07$\\ %Without extinction it is 44.16 L_sun
$6032981^+$  & $ 44 \pm 4$ & $5300 \pm 110$ & -1.01 & 0.37 & $5.188 \pm 0.017$ & $35.4 \pm 0.6$ & $1.15$ & $ 321 \pm 3$ & $0.68 \pm 0.08$\\ %Without extinction it is 48.88 L_sun
8694070  &$ 53 \pm 5$ & $5300 \pm 30$ & -1.44 & 0.25 & $5.135 \pm 0.018$ & $34.6 \pm 0.6$ & $0.7$ & $332 \pm 4$& $0.81 \pm 0.09$\\ \midrule %Without extinction it is 54.41 L_sun
Mock rHB & 44 & 5663  & -1.00 & 0.2 & 6.41& 42.5 & 0.65 & 324 & 0.65\\% 44.07 L_sun
Mock RC & 59 & 4891 & 0.00 & 0.0 & 4.79 & 44.1 & 0.25 & 313 & 1.50\\ \bottomrule % 58.77 L_sun
\end{tabular}
\tablefoot{For each \kepler ID (KIC) we report the effective temperature $T_{\rm eff}$, $\rm [Fe/H]$, $\rm [\alpha/Fe]$ from APOGEE-DR17, or APOGEE-DR16 (one star tagged with a + in apex); mean large frequency separation $\left<\Delta \nu\right>$, and frequency of maximum power $\nu_{\rm max}$ calculated by us using the code in \citet[][]{2016AN....337..774D}, or \citet[][]{2018ApJS..236...42Y} data (one star tagged with a * in apex). The coupling factor $q$ and asymptotic period spacing of the dipole modes $\Delta \Pi_1$ as calculated from the stretched-period method \citep[see e.g.][]{2016A&A...588A..87V}. The current stellar mass $M$ is computed from Eq.~\ref{eq:mass_scaling}. The last two rows show the properties of a simulated rHB and RC star (Sect.~\ref{sec:anal}).}
\end{threeparttable}
}
\end{table*}
%%%%%%%%%%%%%%%%%%%%%%%%%%%%%%%%%%%%%%%%%%%%%%%%%%%%%%%%%%%%%%%%%%%%%%%%%%%%%%%%%%%%%%%%%%%%%%%%

%%%%%%%%%%%%%%%%%%%%%%%%%%%%%%%%%%%%%%%%%%%%%%%%%%%%%%%%%%%%%%%%%%%%%%%%%%%%%%%%%%%%%%%%%%%%%%%%
\section{Simulated data}
%%%%%%%%%%%%%%%%%%%%%%%%%%%%%%%%%%%%%%%%%%%%%%%%%%%%%%%%%%%%%%%%%%%%%%%%%%%%%%%%%%%%%%%%%%%%%%%%
\label{sec:anal}
The aim of this work is not to fit the available observational data, but to analyse the relation between the structures of rHB stars, according to stellar evolution theory, and their oscillation spectra, and to compare the latter with those observed in our sample.  

From a grid of models (see Appendix \ref{app:b}) we selected two sets of parameters that represent well the mass and chemical composition of the classical rHB ($M = 0.65 \ M_\odot$, $\rm [\alpha/Fe] = 0.2$ and $\rm [Fe/H]=-1.00$) and  metal-rich low-mass CHeB ($M = 0.75 \ M_\odot$, $\rm [\alpha/Fe] = 0$ and $\rm [Fe/H]=0$) stars. For comparison, we also consider a typical RC star ($M = 1.5$~\msol with solar composition). As it appears in Fig.~\ref{fig:HR}, the parameters selected for our reference models provide indeed a good representation of the low-intermediate metallicity and metal-rich rHBs in our sample. We also note that without complementary information, such as that provided by asteroseismology, a metal-rich rHB would be mistaken for a more massive star in RGB \citep[see also][]{2017MNRAS.472..979H}.

It is generally accepted that, except for the age, the properties of a low-mass star with a He core of $\simeq0.5$~\msol and an H-rich envelope of $\sim 0.1-0.2$~\msol are largely independent of whether the star was born with a small mass or whether it originates from a more massive star ($M \lesssim 1.8$~\msol) that underwent significant mass loss. Therefore, it is justified to use structure models calculated without mass loss such as those in our grid.

In the following we concentrate on a metal-poor model since, as described in Sect.~\ref{sec:obs}, we expect metal-poor rHBs to present more marked differences with respect to  the spectra of typical RC stars. We select structure models with a central He mass fraction $Y_c \sim 0.27$ %(orange dot in blue track for rHB star, and blue dot in green track for RC target) 
as representative of the CHeB phase. The structures and oscillation spectra of these reference models will be discussed in Sect.~\ref{sec:results}. 

To simulate 4-yr long {\it Kepler} observations of such objects we use the code \texttt{AADG3} \citep[AsteroFLAG Artificial Dataset Generator, version 3.0.2;][and references therein]{2018ApJS..239...34B}. Frequencies and normalised inertiae $E_{\rm norm}$ \citep[see the definition in, e.g.,][]{Aerts2010} of radial ($\ell=0$) and non-radial ($\ell=1-3$) adiabatic oscillation modes are computed using the code \texttt{GYRE} \citep[version 6.0.1,][]{2013MNRAS.435.3406T,2018MNRAS.475..879T,2020ApJ...899..116G}.  \texttt{AADG3} also requires information on modes lifetimes, a quantity directly related to non-adiabatic processes, and therefore not resulting from the \texttt{GYRE} computation.  \texttt{AADG3} uses a relation between $\Gamma_0$, $\nu$, $\nu_{\rm max}$ and \teff calibrated on a small sample of  main sequence and RGB spectra. Since the temperatures of our metal-poor rHBs are outside the domain covered by the calibration sample, and since $\Gamma_0$ also depends on the evolutionary state \citep[][]{Vrard2018}, we adopt as values of $\Gamma_0$ the ones obtained from peak-bagging radial modes in the spectra of our CHeB sample \citep[using the method described in][]{2016AN....337..774D}.

%%%%%%%%%%%%%%%%%%%%%%%%%%%%%%%%%%%%%%%%%%%%%%%%%%%%%%%%%%%%%%%%%%%%%%%%%%%%%%%%%%%%%%%%%%%%%%%%
\section{Discussion}
%%%%%%%%%%%%%%%%%%%%%%%%%%%%%%%%%%%%%%%%%%%%%%%%%%%%%%%%%%%%%%%%%%%%%%%%%%%%%%%%%%%%%%%%%%%%%%%%
\label{sec:results}
In this section we analyse the structures and oscillation spectra of our reference models (rHB and RC), and we compare the simulated PSD with the observed ones (Section \ref{sec:psd}).

%%%%%%%%%%%%%%%%%%%%%%%%%%%%%%%%%%%%%%%%%%%%%%%%%%%%%%%%%%%%%%%%%%%%%%%%%%%%%%%%%%%%%%%%%%%%%%%%
\subsection{Propagation diagram}
\begin{figure*}
\centering
\includegraphics[width=\columnwidth]{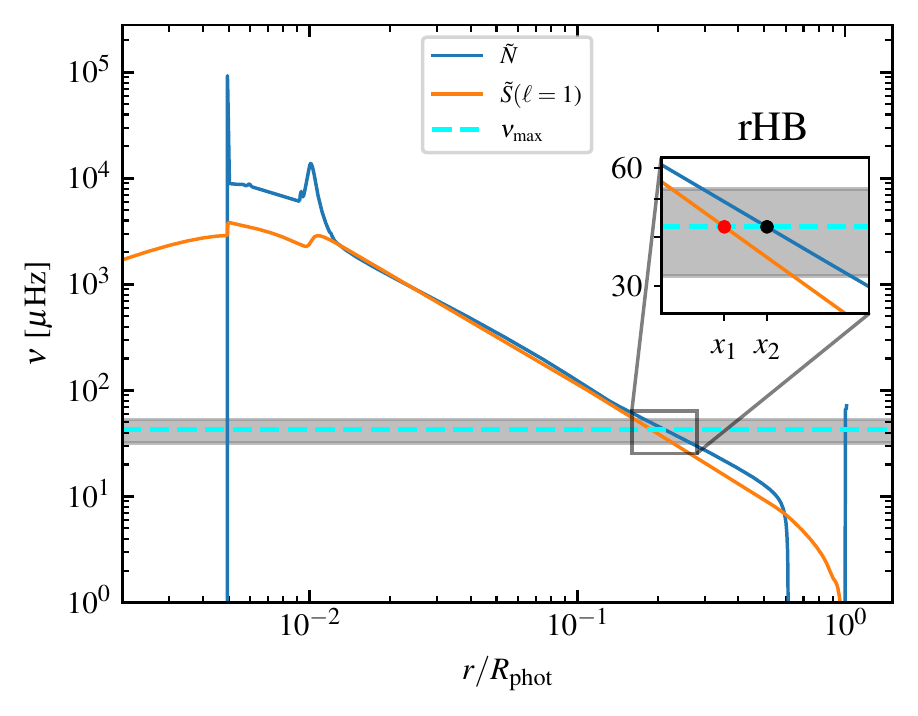}
\includegraphics[width=\columnwidth]{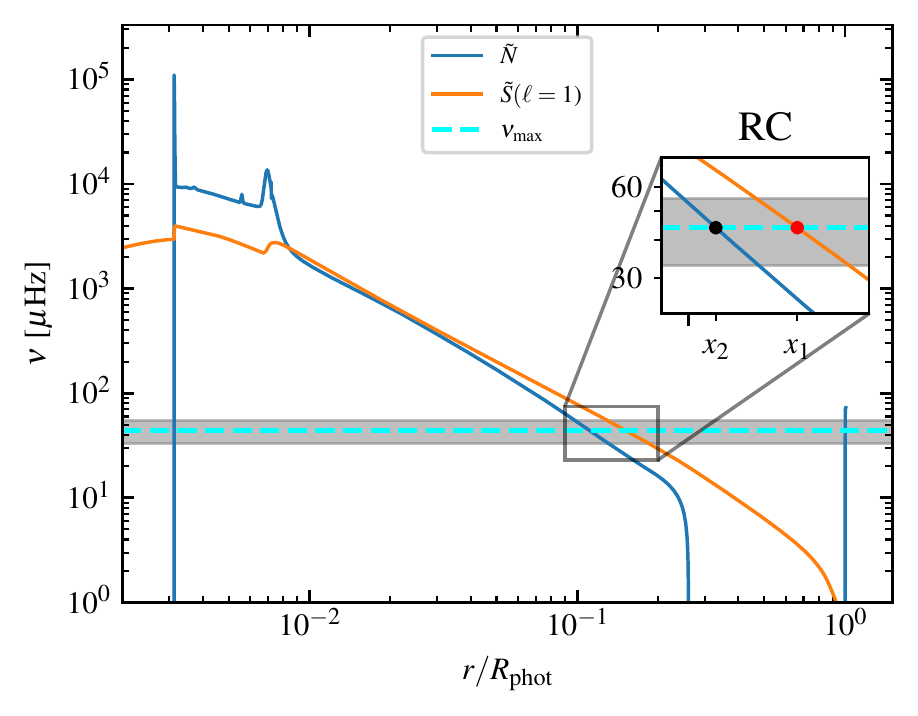}
\includegraphics[width=\columnwidth]{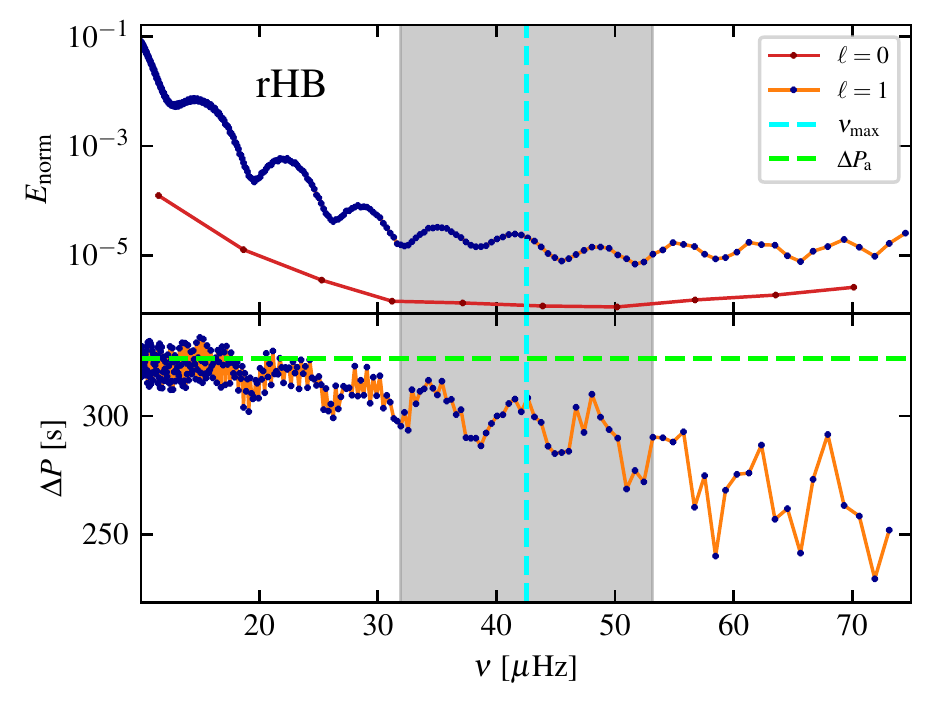}
\includegraphics[width=\columnwidth]{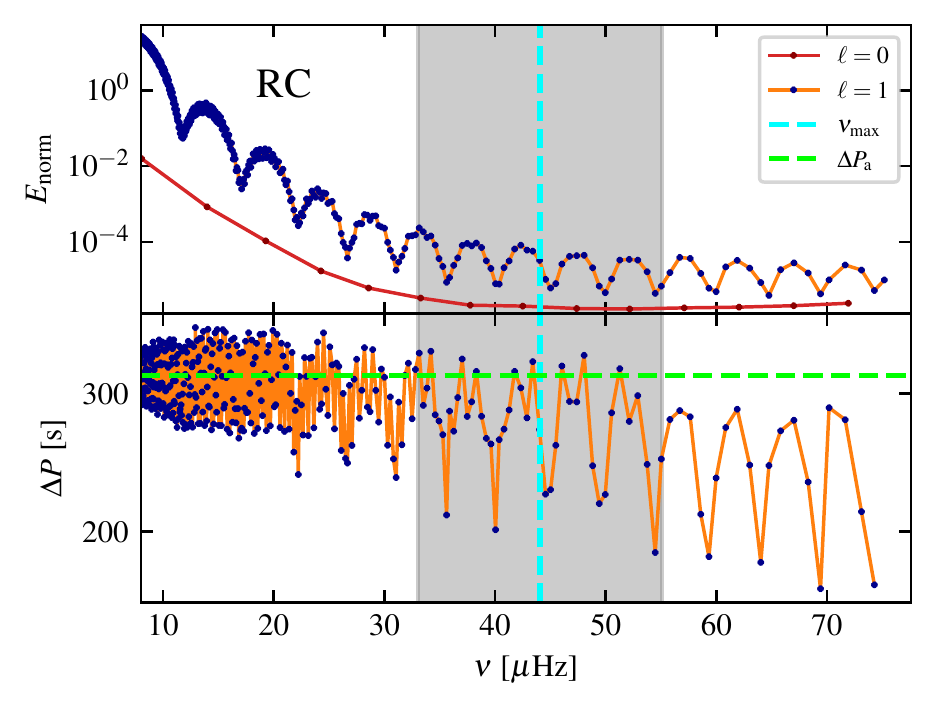}
\caption{Comparison between structure and seismic properties of rHB (left panels) and RC (right panels) reference models (see Sect.~\ref{sec:anal}). {\it Upper panels}:  Propagation diagrams of the dipole modes, with blue and orange lines corresponding to the modified Brunt-V\"ais\"l\"a and Lamb frequencies \citep[][]{2006PASJ...58..893T} respectively. The grey bands represent the frequency domain of expected solar-like oscillations, and at their centre the dashed cyan lines indicate the $\nu_{\rm max}$ values. The insets are zoom-in of the evanescent zones, delimited at $\nu_{\rm max}$ by the red and black points. Their different extension translates in different coupling between g and p cavities (see main text).
{\it Lower panels}: Normalised inertia $E_{\rm norm}$ and period spacing of the dipole modes $\Delta P$ as functions of the eigenfrequencies, with the red curve representing, for comparison, $E_{\rm norm}$ of radial modes. The dashed green line indicates the value of the period spacing from the asymptotic theory of high-order g-modes \citep[$\Delta P_{\rm a}$,][]{Tassoul1980} and the grey band and the dashed cyan line have the same meaning as in the upper panels.}

\label{fig:propdiagramcorrected_l1}
\end{figure*}

%%%%%%%%%%%%%%%%%%%%%%%%%%%%%%%%%%%%%%%%%%%%%%%%%%%%%%%%%%%%%%%%%%%%%%%%%%%%%%%%%%%%%%%%%%%%%%%%
\label{sec:propdiag}
The propagation diagrams of dipole modes for our rHB and RC reference models are shown in the upper panels of Fig.~\ref{fig:propdiagramcorrected_l1}. In each panel, we show the modified Brunt-Väisälä ($\tilde{N}$) and Lamb ($\tilde{S}$) frequencies \citep[][]{2006PASJ...58..893T} as a function of the normalised radius ($x=r/R_{\rm phot}$, with $R_{\rm phot}$ the photospheric radius), as well as the expected frequency domain of the solar-like oscillations. %The latter is represented by the grey band centred at $\nu_{\rm max}$ (dashed cyan line) and extending from  $0.75 \nu_{\rm max}$ to $1.25 \nu_{\rm max}$ \citep[see e.g.][]{2011ApJ...743..161W}.
%%%%%%%%%%%%%%%%%%%%%%%%%%%%%%%%%%%%%%%%%%%%%%%%%%%%%%%%%%%%%%%%%%%%%%%%%%%%%%%%%%%%%%%%%%%%%%%%

The profiles of $\tilde{N}$ and $\tilde{S}$ define the inner limits of the g- and p-cavities. For modes with frequency close to $\nu_{\rm max}$ these limits are defined by the condition $\tilde{S}(x_1) = \nu_{\rm max}$ and $\tilde{N}(x_2) = \nu_{\rm max}$, and in the region between $x_1$ and $x_2$ %(red and black points respectively in the zoom-in boxes) 
the modes are evanescent.

The extent of the evanescent zone is one of the ingredients determining the coupling between resonant cavities \citep[][]{2016PASJ...68..109T,2020A&A...634A..68P}. From the zoom-in boxes in Fig.~\ref{fig:propdiagramcorrected_l1} it appears that this region is smaller in the rHB model than in the RC one and, therefore, we expect the coupling factor $q$ to be larger in the former than in the latter. Indeed, using the structure of our reference models and the strong-coupling approximation\footnote{The weak-coupling approximation \citep[see e.g.][]{1979PASJ...31...87S,1989nos..book.....U} does not hold for low-mass CHeB stars \citep[see e.g.][van Rossem in prep.]{2016A&A...588A..87V,2017A&A...600A...1M}.} for the dipole modes \citep[][]{2016PASJ...68..109T,2020A&A...634A..68P} we obtain $q_{\rm rHB} = 0.65$ and $q_{\rm RC}=0.25$ at $\nu= \nu_{\rm max}$. We note that these values are consistent, given the typical uncertainties ($\sigma_q\sim0.2$), with those % $q$ values determined from the analysis of the
measured from the observed PSDs \citep[see Table~\ref{tab:sample}, \ref{tab:sample_full}, and][]{2016A&A...588A..87V,2017A&A...600A...1M,Mosser2018}.

We notice that the value of the coupling factor is also a function of the mode frequency \citep[e.g.][and van Rossem in prep.]{2020A&A...634A..68P,2020MNRAS.495..621J}.  As shown in Fig.~\ref{fig:propdiagramcorrected_l1}, the size of the evanescent zone decreases (thus $q$ increases) with increasing frequency. The value of $q$ varies from 0.56 to 0.74 in the solar-like frequency domain for the rHB model, and from 0.22 to 0.24 for the RC one. In Sect.~\ref{sec:perspac} we discuss the effect of this variation on the behavior of the period spacing.

%%%%%%%%%%%%%%%%%%%%%%%%%%%%%%%%%%%%%%%%%%%%%%%%%%%%%%%%%%%%%%%%%%%%%%%%%%%%%%%%%%%%%%%%%%%%%%%%
\subsection{Dipole mode properties}
%%%%%%%%%%%%%%%%%%%%%%%%%%%%%%%%%%%%%%%%%%%%%%%%%%%%%%%%%%%%%%%%%%%%%%%%%%%%%%%%%%%%%%%%%%%%%%%%
\label{sec:perspac}
In this section we analyse the properties of the dipole mode spectra computed for our reference models. The bottom panels of Fig.~\ref{fig:propdiagramcorrected_l1} show $E_{\rm norm}$ and the period spacing $\Delta P$ (i.e. period difference between two consecutive modes of same angular degree) as a function of the eigenfrequencies.

We remind that $E_{\rm norm}$ is an average of the mode energy and its value indicates the main region probed by the mode. Modes examining central, high-density regions have higher $E_{\rm norm}$ than modes that are preferentially trapped in the outer regions.
The inertia of dipole modes of the RC model shows a significant variation between local minima and maxima (ratio up to $\approx 27$ in the observable region) corresponding to the p-like and g-like modes, respectively. On the contrary, the inertia in the rHB is almost uniform, with a small contrast between maxima and minima (ratio up to $\approx 3$ in the observable region). This indicates that the dipole modes in the rHB are not clearly trapped in any of the resonant cavities, that is, they have an important mixed p/g character. This behaviour is consistent with the coupling factor values derived in the previous section.% using the strong-coupling approximation \citep{2016PASJ...68..109T} for the reference models.

Since the amplitude of the modes is inversely proportional to the square root of the inertia \citep[see e.g.][]{2009A&A...506...57D}, we expect a modulation of the dipole mode amplitudes around the p-like mode in the case of the RC, as observed in some \kepler red giants, while many dipole modes with similar amplitudes may be observed in the spectrum of the rHB. This implies an increasing complexity of the oscillation spectra, as shown by the stars in our sample (see Fig.~\ref{fig:KICs}).

\begin{figure}
%%%%%%%%%%%%%%%%%%%%%%%%%%%%%%%%%%%%%%%%%%%%%%%%%%%%%%%%%%%%%%%%%%%%%%%%%%%%%%%%%%%%%%%%%%%%%%%%
%%%%%%%%%%%%%%%%%%%%%%%%%%%%%%%%%%%%%%%%%%%%%%%%%%%%%%%%%%%%%%%%%%%%%%%%%%%%%%%%%%%%%%%%%%%%%%%%
\includegraphics[width=\columnwidth]{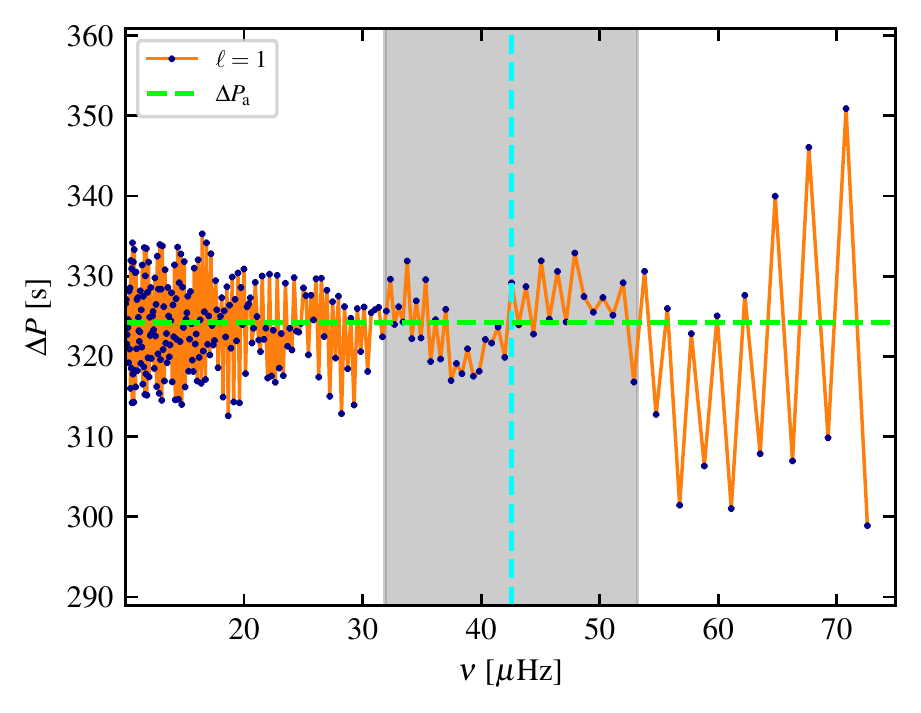}
%%%%%%%%%%%%%%%%%%%%%%%%%%%%%%%%%%%%%%%%%%%%%%%%%%%%%%%%%%%%%%%%%%%%%%%%%%%%%%%%%%%%%%%%%%%%%%%%
\caption{Period spacing as a function of the eigenfrequencies of the isolated dipole $\gamma$-modes \citep[][]{2020ApJ...898..127O}. Other symbols and colors are the same as in Fig.~\ref{fig:propdiagramcorrected_l1}. The high modulation in the period spacing above the observable frequencies is connected to structural glitches \citep[see e.g.][]{2015MNRAS.453.2290B}.}
\label{fig:profile_periodspacing_l1_gamma2}
\end{figure}

The high value of $q$ also affects the behaviour of the period spacing \citep[see also][]{2017A&A...600A...1M}. In the bottom part of the lower panels of Fig.~\ref{fig:propdiagramcorrected_l1}, we plot $\Delta P$ as a function of the eigenfrequencies as well as the constant value (green dashed line) predicted by the asymptotic g-mode approximation \citep[$\Delta P_{\rm a}$,][]{Tassoul1980}. In the observable frequency domain  we notice for the rHB model a significant deviation of $\Delta P$ from the asymptotic value even for modes with high inertia, as well as a decreasing trend of $\Delta P$ with increasing frequency.
To show that both effects are a consequence of the high value of $q$ and its frequency dependence, we use the \citet[][]{2020ApJ...898..127O} formalism to separate pure isolated p-modes ($\pi$-modes) from pure isolated g-modes ($\gamma$-modes), that is, pure g-modes not affected by the coupling with the acoustic cavity. In Fig.~\ref{fig:profile_periodspacing_l1_gamma2} we plot the period spacing of dipole $\gamma$-modes, and, as we could expect, their average value is consistent with that from the asymptotic approximation of pure high-order g-modes. 
Therefore, the differences in the period spacing of the RC and rHB models are explained by the high coupling for the latter, which causes all dipole modes to have an important acoustic component, thus decreasing the value of $\Delta P$.

%%%%%%%%%%%%%%%%%%%%%%%%%%%%%%%%%%%%%%%%%%%%%%%%%%%%%%%%%%%%%%%%%%%%%%%%%%%%%%%%%%%%%%%%%%%%%%%%
%%%%%%%%%%%%%%%%%%%%%%%%%%%%%%%%%%%%%%%%%%%%%%%%%%%%%%%%%%%%%%%%%%%%%%%%%%%%%%%%%%%%%%%%%%%%%%%%

%%%%%%%%%%%%%%%%%%%%%%%%%%%%%%%%%%%%%%%%%%%%%%%%%%%%%%%%%%%%%%%%%%%%%%%%%%%%%%%%%%%%%%%%%%%%%%%%
\subsection{Power spectral density}
\begin{figure*}
%%%%%%%%%%%%%%%%%%%%%%%%%%%%%%%%%%%%%%%%%%%%%%%%%%%%%%%%%%%%%%%%%%%%%%%%%%%%%%%%%%%%%%%%%%%%%%%%
\centering
\includegraphics[width=\columnwidth]{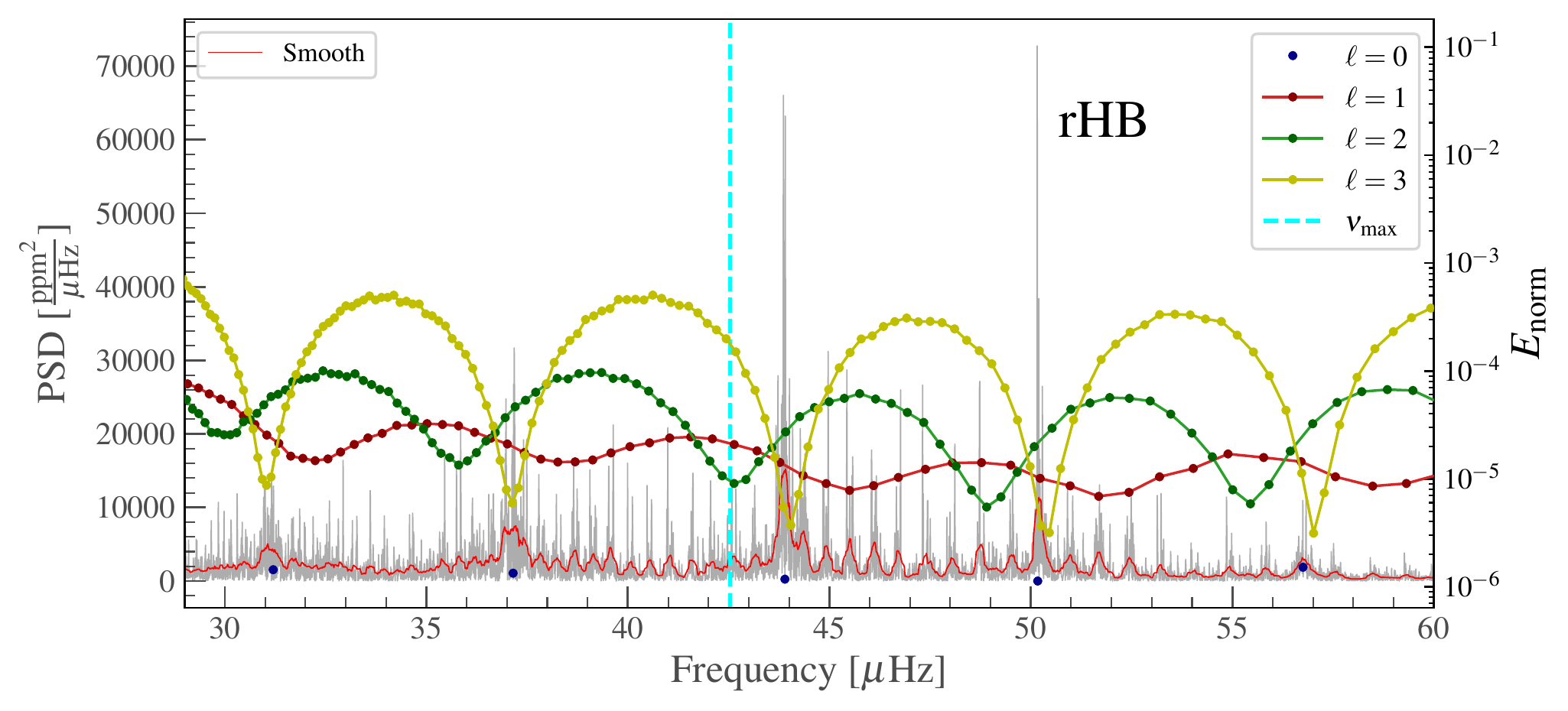}
\includegraphics[width=\columnwidth]{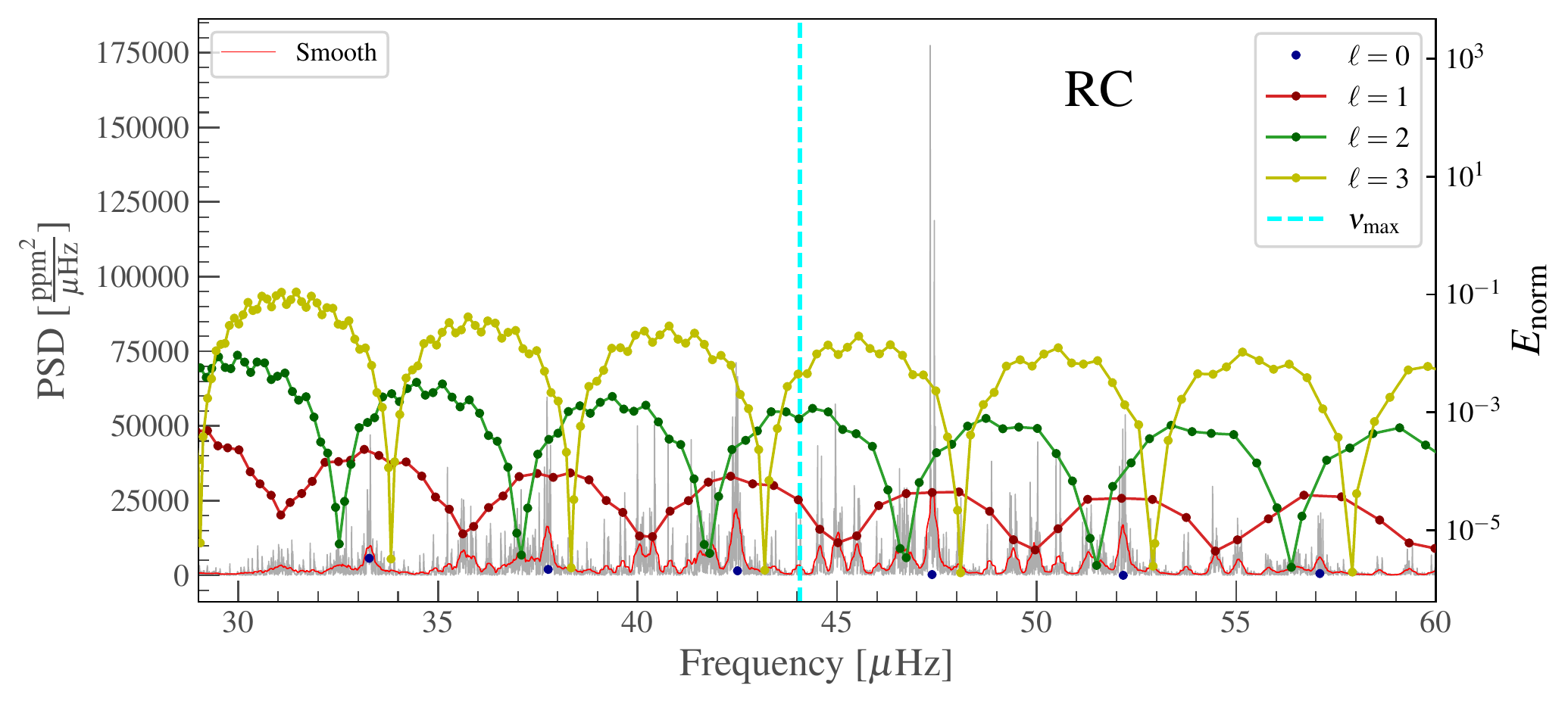}
%%%%%%%%%%%%%%%%%%%%%%%%%%%%%%%%%%%%%%%%%%%%%%%%%%%%%%%%%%%%%%%%%%%%%%%%%%%%%%%%%%%%%%%%%%%%%%%%
\caption{Simulated PSD (grey line) as a function of the eigenfrequencies using theoretical $\ell =0,1,2,3$ modes for rHB (left) and RC (right) models. The red line is a smoothed version of the PSD. The vertical dashed cyan lines correspond to $\nu_{\rm max}$ values and colored dots and lines  represent the values of $E_{\rm norm}$ for the $\ell =0,1,2,3$ modes.}
\label{fig:eigen_inertia_psd}
\end{figure*}

%%%%%%%%%%%%%%%%%%%%%%%%%%%%%%%%%%%%%%%%%%%%%%%%%%%%%%%%%%%%%%%%%%%%%%%%%%%%%%%%%%%%%%%%%%%%%%%%
\label{sec:psd}
%Using the properties of the theoretical spectra for rHB and RC models and the code \texttt{AADG3}, we simulate 4~yr of {\it Kepler} observations for these targets. 
Fig.~\ref{fig:eigen_inertia_psd} shows the simulated PSDs of our reference models %their PSD around \numax, 
together with the inertia of the $\ell =0,1,2,3$ modes. The contribution of each degree to the PSD is shown in Appendix~\ref{app:c}.

Comparison between Fig.~\ref{fig:eigen_inertia_psd} and Fig.~\ref{fig:KICs} shows many similarities between the rHB-mock spectrum and the observed ones. These spectra appear noisier than RC ones, with a large number of peaks corresponding to non-radial modes. In particular, there are observable dipole modes in the entire frequency range between two consecutive radial modes, unlike the behaviour in RC and low-RGB stars, where only a few modes around the corresponding p-like mode have observable amplitudes. 

We see that the strong coupling also affects the quadrupole modes. Several of them, with frequencies close to those of the p-like modes, are expected to have similar contributions to the PSD. Moreover, because of a higher inertia at the local minima with respect to the RC model, quadrupole modes in rHB stars would have lower amplitudes. All that makes more challenging to detect and characterise $\ell=2$ modes in CHeB metal-poor stars. 
Finally, $\ell=3$ modes have eigenfrequencies close to those of radial modes and their heights are similar to the background noise. They tend to form a continuum that should be considered during the background analysis (see Appendix~\ref{app:c}).

%%%%%%%%%%%%%%%%%%%%%%%%%%%%%%%%%%%%%%%%%%%%%%%%%%%%%%%%%%%%%%%%%%%%%%%%%%%%%%%%%%%%%%%%%%%%%%%%
%%%%%%%%%%%%%%%%%%%%%%%%%%%%%%%%%%%%%%%%%%%%%%%%%%%%%%%%%%%%%%%%%%%%%%%%%%%%%%%%%%%%%%%%%%%%%%%%

%%%%%%%%%%%%%%%%%%%%%%%%%%%%%%%%%%%%%%%%%%%%%%%%%%%%%%%%%%%%%%%%%%%%%%%%%%%%%%%%%%%%%%%%%%%%%%%%
\section{Conclusions}
%%%%%%%%%%%%%%%%%%%%%%%%%%%%%%%%%%%%%%%%%%%%%%%%%%%%%%%%%%%%%%%%%%%%%%%%%%%%%%%%%%%%%%%%%%%%%%%%
\label{sec:concl}
High-quality spectra obtained from the 4-year long \kepler observations of a large number of red giants allowed us to identify a small number of red giants (12) whose oscillation spectra appear to be very noisy or complex with respect to the typical behaviour of oscillation spectra in \kepler red giants. Their global seismic parameters are compatible with low-mass stars ($M \lesssim 0.8$~\msol) in the central He-burning phase, and the fit of the asymptotic relation for the dipole modes \citep[e.g.][]{2016A&A...588A..87V} results in coupling factor values $q \gtrsim 0.4$, i.e. much higher than the typical value for stars classified as RC \citep[$q\sim 0.25-0.30$, e.g.][]{2016A&A...588A..87V,2017A&A...600A...1M,Mosser2018}. In our sample we find stars with low/intermediate metallicity (75\%), and also with solar metallicity. Their position in the HRD is compatible with the so-called rHB stars, i.e. low-mass objects between the RRL-IS and the RC at the corresponding metallicity. Stellar evolution theory predicts for these stars a structure consisting of a He core of $\sim 0.5$~\msol and an envelope of $\approx 0.1-0.2$~\msol \citep[e.g.][]{1989upsf.conf..103R,2008A&A...487..185V,2010A&A...517A..81G,2016ARA&A..54...95G,2020MNRAS.498.5745T}.

In this work we have shown that the oscillation spectra we expect for this type of stars are entirely consistent with those observed in our sample. These spectra are clearly different from those of the stars that, with a similar He core but a much larger envelope, populate the RC. The main factor determining these differences is the coupling between the inner and outer regions, reflecting very different density profiles inside these stars.
A second factor increasing the complexity of these spectra is the higher temperature of the less metallic stars, which decrease the lifetime of the modes. In fact, solar-like oscillations in rHBs have also been detected in the K2 \citep{2014PASP..126..398H} light curves of the globular cluster M4 \citep[e.g.][]{2019ApJS..244...12W}, where the complexity of the spectra and the reduced observation time (80~days) have made it difficult to extract robust $\langle\Delta\nu\rangle$ values \citep[e.g.][]{2022A&A...662L...7T,2022MNRAS.515.3184H}.

rHB stars are well known and easily identified in globular clusters. Here we have also shown the ability of asteroseismology to identify these low-mass CHeB stars in the field and in solar-metallicity environments where, even with high-precision photometry, they would be hardly distinguishable from other stars in RC or RGB phases. 

It is clear that 0.7~\msol stars, especially those of solar metallicity, must have followed a non-standard evolution during which they have lost a large amount of mass \citep[see also][]{2022NatAs...6..673L,2022arXiv220804332B}. This work provides us with a solid framework for the future study of these stars and of the processes that have led them to their current mass. Knowing that is fundamental in order to derive their ages with accuracy, and to potentially provide another piece of the puzzle in the sequence between RC and subdwarf B stars or other stripped stars.

%%%%%%%%%%%%%%%%%%%%%%%%%%%%%%%%%%%%%%%%%%%%%%%%%%%%%%%%%%%%%%%%%%%%%%%%%%%%%%%%%%%%%%%%%%%%%%%%

\begin{acknowledgements}
We are grateful to (in alphabetical order) Emma Willett, Joel Ong, Joris De Ridder, Masao Takata and Saniya Khan for useful discussions. We are also grateful to the anonymous referee for the constructive comments. This work has made use of data from the European Space Agency (ESA) mission {\it Gaia} (\href{https://www.cosmos.esa.int/gaia}{https://www.cosmos.esa.int/gaia}) and from the Two Micron All Sky Survey (\href{https://irsa.ipac.caltech.edu/Missions/2mass.html}{https://irsa.ipac.caltech.edu/Missions/2mass.html}). This research made use of Lightkurve, a Python package for Kepler and TESS data analysis \citep[][]{2018ascl.soft12013L}, and of dustmaps, a package for  interstellar dust reddening and extinction \citep{2018JOSS....3..695M}. AM, AS, GC, JM, MM, MT acknowledge support from the ERC Consolidator Grant funding scheme (project ASTEROCHRONOMETRY, \href{https://www.asterochronometry.eu}{https://www.asterochronometry.eu}, G.A. n. 772293). MV acknowledge support from NASA grant 80NSSC18K1582. Funding for the Stellar Astrophysics Centre is provided by The Danish National Research Foundation (Grant agreement No.~DNRF106).
\end{acknowledgements}

%%%%%%%%%%%%%%%%%%%%%%%%%%%%%%%%%%%%%%%%%%%%%%%%%%%%%%%%%%%%%%%%%%%%%%%%%%%%%%%%%%%%%%%%%%%%%%%%
\bibliographystyle{aa}
\bibliography{references_rhb}
%%%%%%%%%%%%%%%%%%%%%%%%%%%%%%%%%%%%%%%%%%%%%%%%%%%%%%%%%%%%%%%%%%%%%%%%%%%%%%%%%%%%%%%%%%%%%%%%

\begin{appendix}

%%%%%%%%%%%%%%%%%%%%%%%%%%%%%%%%%%%%%%%%%%%%%%%%%%%%%%%%%%%%%%%%%%%%%%%%%%%%%%%%%%%%%%%%%%%%%%%%
\section{Physical properties of the full sample}
%%%%%%%%%%%%%%%%%%%%%%%%%%%%%%%%%%%%%%%%%%%%%%%%%%%%%%%%%%%%%%%%%%%%%%%%%%%%%%%%%%%%%%%%%%%%%%%%
\label{app:a1}
%%%%%%%%%%%%%%%%%%%%%%%%%%%%%%%%%%%%%%%%%%%%%%%%%%%%%%%%%%%%%%%%%%%%%%%%%%%%%%%%%%%%%%%%%%%%%%%%

\begin{figure}
%%%%%%%%%%%%%%%%%%%%%%%%%%%%%%%%%%%%%%%%%%%%%%%%%%%%%%%%%%%%%%%%%%%%%%%%%%%%%%%%%%%%%%%%%%%%%%%%
\centering
\includegraphics[width=\columnwidth]{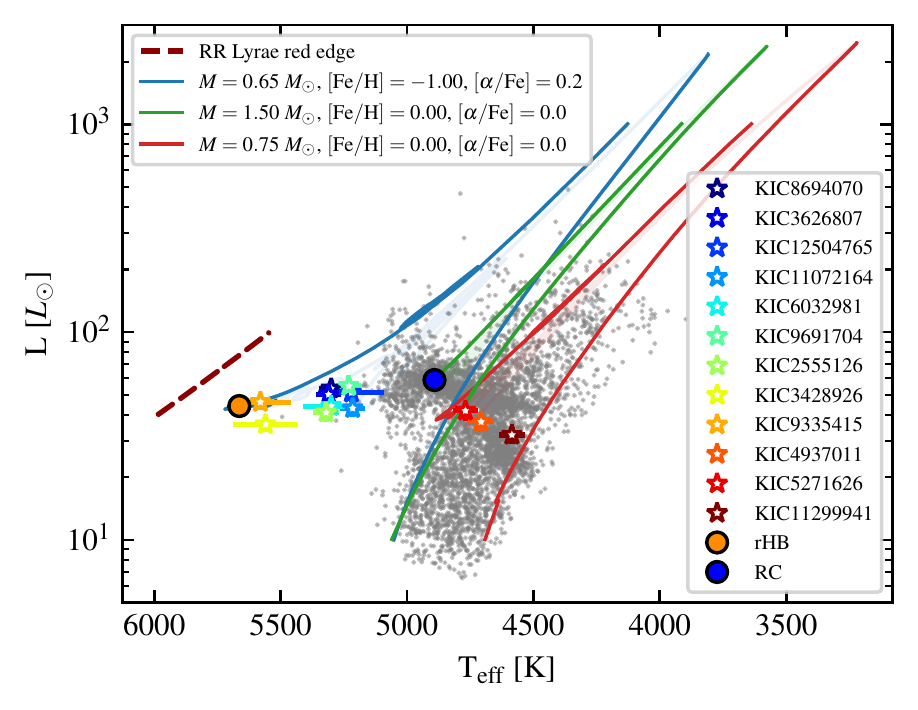}
\caption{The same as Fig.~\ref{fig:HR}, but including all the CHeB stars in our sample. These stars are colour-coded according to increasing [Fe/H].}
\label{fig:full_sample}
\end{figure}

In this appendix we give some details concerning the origin of the physical quantities in Tables~\ref{tab:sample} and ~\ref{tab:sample_full}. The latter complements the former providing the properties of the rest of our sample of rHB candidates (see also the  HRD of the whole sample in Fig.~\ref{fig:full_sample}).
\begin{table*}[hbt!]
\centering
\resizebox{\textwidth}{!}{%
%%%%%%%%%%%%%%%%%%%%%%%%%%%%%%%%%%%%%%%%%%%%%%%%%%%%%%%%%%%%%%%%%%%%%%%%%%%%%%%%%%%%%%%%%%%%%%%%
\begin{threeparttable}
\centering
%%%%%%%%%%%%%%%%%%%%%%%%%%%%%%%%%%%%%%%%%%%%%%%%%%%%%%%%%%%%%%%%%%%%%%%%%%%%%%%%%%%%%%%%%%%%%%%%
%\caption{Physical properties for the rest of our sample of rHB candidates as well as those of KIC4937011 (undermassive star in NGC~6819, tagged with a $R$ in apex), for which we show $\left<\Delta \nu\right>$, \numax, and $M$ from \citet{2017MNRAS.472..979H}. See Table~\ref{tab:sample} for a description of the symbols.}
\caption{Physical properties for the rest of our sample of rHB candidates.}
\label{tab:sample_full}
%%%%%%%%%%%%%%%%%%%%%%%%%%%%%%%%%%%%%%%%%%%%%%%%%%%%%%%%%%%%%%%%%%%%%%%%%%%%%%%%%%%%%%%%%%%%%%%%
\begin{tabular}{@{}lcllllllll@{}}
\toprule
KIC & $L$ [$L_\odot$] & $T_{\rm eff}$ [K]  &  $\rm [Fe/H]$  & $\rm [\alpha/Fe]$  &  $\left<\Delta \nu\right>$ [$\mu$Hz]& $\nu_{\rm max}$ [$\mu$Hz]& $q$ & $\Delta \Pi_1$ [s] & $M$ [$M_\odot$]\\ \midrule
2555126 & $41 \pm 4 $ & $5320 \pm 20$& -0.72 & 0.26 & $5.66 \pm 0.03$ & $36.4 \pm 0.6$ & 0.93  & $ 280 \pm 20$ & $ 0.64\pm 0.06$\\
$3428926^+$ & $36 \pm 3$ & $5560\pm 130$ & -0.50 & 0.27 & $6.72 \pm 0.02$ & $43.0 \pm 0.6$ & 1.15 & $ 270 \pm 40$ & $ 0.58\pm 0.07$\\ 
3626807 & $50 \pm 6$ & $5310 \pm 20$ & -1.16 & 0.26 & $5.276 \pm 0.011$ & $36.5 \pm 0.6$ & 0.69 & $ 308 \pm 6$ & $ 0.79 \pm 0.10$\\
%3759696 & $33 \pm 4$ & $4551 \pm 9$ & 0.27 & 0.06 & $ 4.036 \pm $ & $ 27.23 \pm $ & 0.35 & $ 290 \pm 2$ & $ \pm $\\
%$8299794^+$ & $46 \pm 4$ & $5610 \pm 150$ & -0.69 & 0.28 & $6.168 \pm 0.016$ & $41.0 \pm 0.7$ & 0.05 & $290  \pm 50$ & $ \pm $\\
$9335415^+$ & $46 \pm 4 $ & $5580 \pm 120$ & -0.50 & 0.11 & $5.808 \pm 0.018$ & $34.9 \pm 0.5$ & 0.53 & $ 240  \pm 40$ & $ 0.59 \pm 0.07$\\
9691704 & $55 \pm 7$ & $5230 \pm 20$ & -0.88 & 0.30 & $4.802 \pm 0.013$ & $32.6 \pm 0.5$ & 0.23 & $ 334  \pm 5$ & $ 0.83 \pm 0.11$\\
11072164 & $43 \pm 4$ & $5215 \pm 18$ & -1.01 & 0.24 & $4.761 \pm 0.012$ & $32.8 \pm 0.5$ & 1.11 & $ 300 \pm 50$ & $ 0.65 \pm 0.06$\\ 
$11299941^*$ & $32 \pm 3$ & $4585 \pm 7$ & 0.25 & 0.05 & $ 4.08 \pm 0.09$ & $28.0 \pm 0.8$ & 0.45 & $ 300 \pm 20$ & $0.64 \pm 0.08$\\
$12504765^+$ & $51 \pm 5$ & $5220 \pm 130$ & -1.15 & 0.33 & $4.817 \pm 0.010$ & $32.4 \pm 0.5$ & 0.65  & $ 340 \pm 20$ & $0.76 \pm 0.10$\\
$4937011^R$ & $37 \pm 4$ & $4707\pm 8$ & -0.02 & 0.03 & $4.08 \pm 0.10$ & $28.3 \pm 0.4$ & $0.53$ & $ 224.3 \pm 1.4$ & $0.71 \pm 0.08$\\ \bottomrule% considerando la L verrebbe 0.69 +- 0.08 M_sun
\end{tabular}
\tablefoot{It includes also the properties of KIC4937011 (undermassive star in NGC~6819, tagged with a $R$ in apex), for which we show $\left<\Delta \nu\right>$, \numax, and $M$ from \citet{2017MNRAS.472..979H}. See Table~\ref{tab:sample} for a description of the symbols}
%%%%%%%%%%%%%%%%%%%%%%%%%%%%%%%%%%%%%%%%%%%%%%%%%%%%%%%%%%%%%%%%%%%%%%%%%%%%%%%%%%%%%%%%%%%%%%%%
%\begin{tablenotes}
%\item[a] \label{foot:a} \citet[][]{2022ApJS..259...35A};
%\item[b] \label{foot:b} \citet[][]{2020ApJS..249....3A};
%\item[c] \label{foot:c} \citet[][]{2019arXiv190609428K};
%\item[d] \label{foot:d} \citet[][]{2009A&A...508..877M,2011A&A...532A..86M}
%\end{tablenotes}
%%%%%%%%%%%%%%%%%%%%%%%%%%%%%%%%%%%%%%%%%%%%%%%%%%%%%%%%%%%%%%%%%%%%%%%%%%%%%%%%%%%%%%%%%%%%%%%%
\end{threeparttable}
}
\end{table*}

The global seismic parameters \numax and $\langle\Delta\nu\rangle$ %of the most metal rich targets 
of targets tagged with an asterisk in Tables~\ref{tab:sample} and ~\ref{tab:sample_full} are taken from \citet{2018ApJS..236...42Y}, while those for the NGC~6819 cluster member (KIC~4937011, tagged with $^R$) are from \citet[][]{2017MNRAS.472..979H}.
For the rest of the sample, we employ the approach of \citet[][]{2016AN....337..774D} and the value of $\langle\Delta\nu\rangle$ is computed using individual frequencies and the weighted fit of the asymptotic relation for radial modes. As discussed in \citet[][]{2017MNRAS.472..979H}, this method gives results in good agreement with the values of $\langle\Delta\nu\rangle$ derived by \citet{2018ApJS..236...42Y} and allows a forward comparison with model-based values. The asymptotic period spacing of the dipole modes $\Delta \Pi_1$ and the coupling factor $q$ are derived using the stretched-period method \citep[see e.g.][]{2016A&A...588A..87V}.

The atmospheric parameters \teff and chemical composition come from APOGEE-DR17, except for four targets with a \texttt{STAR\_BAD} flag in that release. For them ('+' apex in Table~\ref{tab:sample} and \ref{tab:sample_full}) we adopt the available values in APOGEE-DR16. To check the reliability of these atmospheric parameters and of the quoted uncertainties, we performed an independent analysis for the  three stars in Table~\ref{tab:sample}. We used \texttt{MOOG-synth}\footnote{\href{https://www.as.utexas.edu/~chris/moog.html}{https://www.as.utexas.edu/~chris/moog.html}} with the assumption of local thermodynamic equilibrium, the linelist of APOGEE-DR17 \citep{2015ApJS..221...24S,2021AJ....161..254S} implemented with lines from \texttt{VALD} database\footnote{\href{http://vald.astro.uu.se}{http://vald.astro.uu.se}} and \texttt{MARCS} model atmospheres \citep{2008A&A...486..951G}. We get results in good agreement with those in APOGEE-DR16/17 except for \teff uncertainties. Even for the best situation in which $\log \, g$ is fixed to the seismic values \citep[e.g.][]{2019A&A...627A.173V}, the uncertainty on \teff is $\sigma_{T_{\rm eff}}\sim 50$~K. Therefore, although in Tables~\ref{tab:sample} and \ref{tab:sample_full} we keep the values from APOGEE, we assume a minimum value of $\sigma_{\teff}=50$~K in deriving stellar mass and its uncertainty.

Bolometric luminosities $L$  are estimated by combining astrometry data from {\it Gaia} DR3 \citep[][]{2022arXiv220605989B} with 2MASS photometry \citep[][]{2006AJ....131.1163S} in the $\rm K_s$-band and bolometric correction from \citet[][]{2014MNRAS.444..392C,2018MNRAS.479L.102C}.
We applied the {\it Gaia}-DR3 parallax zero-point correction of \citet[][]{2021A&A...649A...4L} and estimated reddening and extinction from the three-dimensional maps of \citet[][]{2019ApJ...887...93G}. The errors in $L$ are calculated with a Markov chain Monte Carlo (MCMC) method considering fixed the extinction and the value of $\rm M_{bol,\odot}$ \citep[$\rm M_{bol,\odot} = 4.75$, ][]{2014MNRAS.444..392C}.

Stellar masses, as described in Sec.~\ref{sec:obs}, have been estimated using the scaling relation Eq.~\ref{eq:mass_scaling} and the values of $L$, \teff and \numax just described. In the following paragraph we present the results obtained with an alternative scaling relation. 

\subsection{Stellar mass from scaling relation involving $\langle \Delta\nu\rangle$ and \numax}
%%%%%%%%%%%%%%%%%%%%%%%%%%%%%%%%%%%%%%%%%%%%%%%%%%%%%%%%%%%%%%%%%%%%%%%%%%%%%%%%%%%%%%%%%%%%%%%%
\label{app:a2}
In order to test the mass estimations made with Eq.~\ref{eq:mass_scaling} of Sect.~\ref{sec:obs}, we employed the model-based corrected scaling relation \citep[see e.g.][]{1995A&A...293...87K,2011ApJ...730...63G}
%%%%%%%%%%%%%%%%%%%%%%%%%%%%%%%%%%%%%%%%%%%%%%%%%%%%%%%%%%%%%%%%%%%%%%%%%%%%%%%%%%%%%%%%%%%%%%%%
\begin{equation}
\label{eq:mass_scaling_appendix}
\centering
    \frac{M}{M_\odot} = f_{\Delta \nu}^4 \ \left(\frac{\rm T_{eff}}{\rm T_{eff,\odot}}\right)^{1.5}\left(\frac{\nu_{\rm max}}{\nu_{\rm max,\odot}}\right)^3 \left(\frac{\left<\Delta \nu\right>_\odot}{\left<\Delta \nu\right>}\right)^4
\end{equation}
%%%%%%%%%%%%%%%%%%%%%%%%%%%%%%%%%%%%%%%%%%%%%%%%%%%%%%%%%%%%%%%%%%%%%%%%%%%%%%%%%%%%%%%%%%%%%%%%
for two metal-rich stars (KIC5271626 and KIC4937011) and for two metal-poor stars (KIC6032981 and KIC11072164) of our sample. Here we used the solar reference values of Sect.~\ref{sec:obs}, and $\left<\Delta \nu\right>_\odot = 135.1 \ \mu$Hz \citep[][]{2011ApJ...743..143H}. The correction factor $f_{\Delta \nu}$ on the $\langle \Delta\nu\rangle$ scaling law \citep{Ulrich1986} is derived with the procedure described in \cite{2017MNRAS.467.1433R}, i.e. by using the theoretical radial mode frequencies of stellar models to compute $\langle \Delta\nu\rangle$ from the weighted linear fit of the asymptotic relation \citep[see also][]{2021A&A...645A..85M, 2022A&A...662L...7T}. We based the iterative search for the correct $f_{\Delta \nu}$ on evolutionary tracks with the same metallicity (within the errors) as the four stars aforementioned: solar composition for the metal-rich ones; $\rm [Fe/H] =-1.00$ with $\rm [\alpha/Fe] =0.2$ and $\rm [\alpha/Fe] =0.4$ for the two metal-poor ones (see Appendix \ref{app:b} for details on the models). To correct the model-predicted $\langle \Delta\nu\rangle$ from the surface effects, we included $\left<\Delta \nu\right>_\odot = 135.3 \ \mu$Hz of our solar-calibrated model to the correction factor $f_{\Delta \nu}$ \citep[e.g.][]{2011ApJ...743..161W}. Finally, we computed the theoretical radial oscillations with the tool \texttt{GYRE}.
%We also use the theoretical radial mode frequencies of stellar models to compute $\langle \Delta\nu\rangle$ from the weighted linear fit of the asymptotic relation \citep[see e.g.][]{2017MNRAS.467.1433R} and estimate the correction factor $f_{\Delta \nu}$ in Eq.~\ref{eq:mass_scaling_appendix} \citep[see also][]{2017MNRAS.467.1433R, 2021A&A...645A..85M, 2022A&A...662L...7T}
The $f_{\Delta \nu}$ we found are nearly equal to 1.03 and 1.01 for the metal-poor and for the metal-rich stars respectively. In deriving the masses with Eq.~\ref{eq:mass_scaling_appendix}, we considered a minimum error of 50 K in \teff (as said previously in Appendix \ref{app:a1}), and an error of 0.01 on $f_{\Delta \nu}$ due to the impossibility of knowing the exact position, at fixed \numax, of our observed stars along the evolutionary tracks. Therefore, these masses are compatible within the errors with those derived from Eq.~\ref{eq:mass_scaling}. We want also to notice that it is difficult to have a very precise $\left<\Delta \nu\right>$ estimate for these stars, because the radial modes are located in crowded regions (see Appendix \ref{app:c}). This leads to systematic errors in the measurement of individual radial modes that can be of the order of 4\% by mass.
\FloatBarrier

%%%%%%%%%%%%%%%%%%%%%%%%%%%%%%%%%%%%%%%%%%%%%%%%%%%%%%%%%%%%%%%%%%%%%%%%%%%%%%%%%%%%%%%%%%%%%%%%
\section{Grids of stellar models}
%%%%%%%%%%%%%%%%%%%%%%%%%%%%%%%%%%%%%%%%%%%%%%%%%%%%%%%%%%%%%%%%%%%%%%%%%%%%%%%%%%%%%%%%%%%%%%%%
\label{app:b}
As mentioned in Sect.~\ref{sec:anal}, we chose three sets of stellar parameters  to represent a rHB star, a metal-rich low-mass CHeB star, and a RC star. The stellar models at the base of this work belong to a grid of stellar evolutionary models computed with the code \texttt{MESA-r11532} \citep[Modules for Experiments in Stellar Astrophysics][] {2011ApJS..192....3P,2013ApJS..208....4P,2015ApJS..220...15P,2016ApJS..223...18P,2018ApJS..234...34P,2019ApJS..243...10P}. In the computation we follow the evolution from the pre-main sequence phase until the first thermal pulse  in the asymptotic giant branch  for stellar masses from 0.6~\msol till 2.00~\msol, with a step of 0.05~\msol. We consider 36 different chemical composition, with 12 values of [Fe/H] (from -2.5 to 0.25) and  three values of alpha-elements enhancement: $\rm[\alpha/Fe]=0.0$, 0.2 and 0.4. We adopt as reference solar mixture that from \cite{AGSS09} and high- and low-temperature radiative opacity tables have been computed for these specific metal mixtures,  solar and alpha-enhanced ones.  Envelope convection is described by the Mixing Length theory \cite{CoxGiuli1968} and the corresponding $\alpha_{\rm MLT}$ parameter, the same for all the grid, is derived from the solar calibration with the same physics. We add below the convective envelope a diffusive undershooting \citep{Herwig2000} with a size parameter $f=0.02$ \citep[see][]{Khan2018}. Extra-mixing over the convective core limit during the central-He burning phase is treated following the formalism by \cite{2017MNRAS.469.4718B}.

%%%%%%%%%%%%%%%%%%%%%%%%%%%%%%%%%%%%%%%%%%%%%%%%%%%%%%%%%%%%%%%%%%%%%%%%%%%%%%%%%%%%%%%%%%%%%%%%
\section{Contribution of individual eigenmodes to the PSDs of CHeB stars}
%%%%%%%%%%%%%%%%%%%%%%%%%%%%%%%%%%%%%%%%%%%%%%%%%%%%%%%%%%%%%%%%%%%%%%%%%%%%%%%%%%%%%%%%%%%%%%%%
\label{app:c}

\begin{figure}[hb]
%%%%%%%%%%%%%%%%%%%%%%%%%%%%%%%%%%%%%%%%%%%%%%%%%%%%%%%%%%%%%%%%%%%%%%%%%%%%%%%%%%%%%%%%%%%%%%%%
\centering
\vspace*{1cm}
\includegraphics[width=\columnwidth]{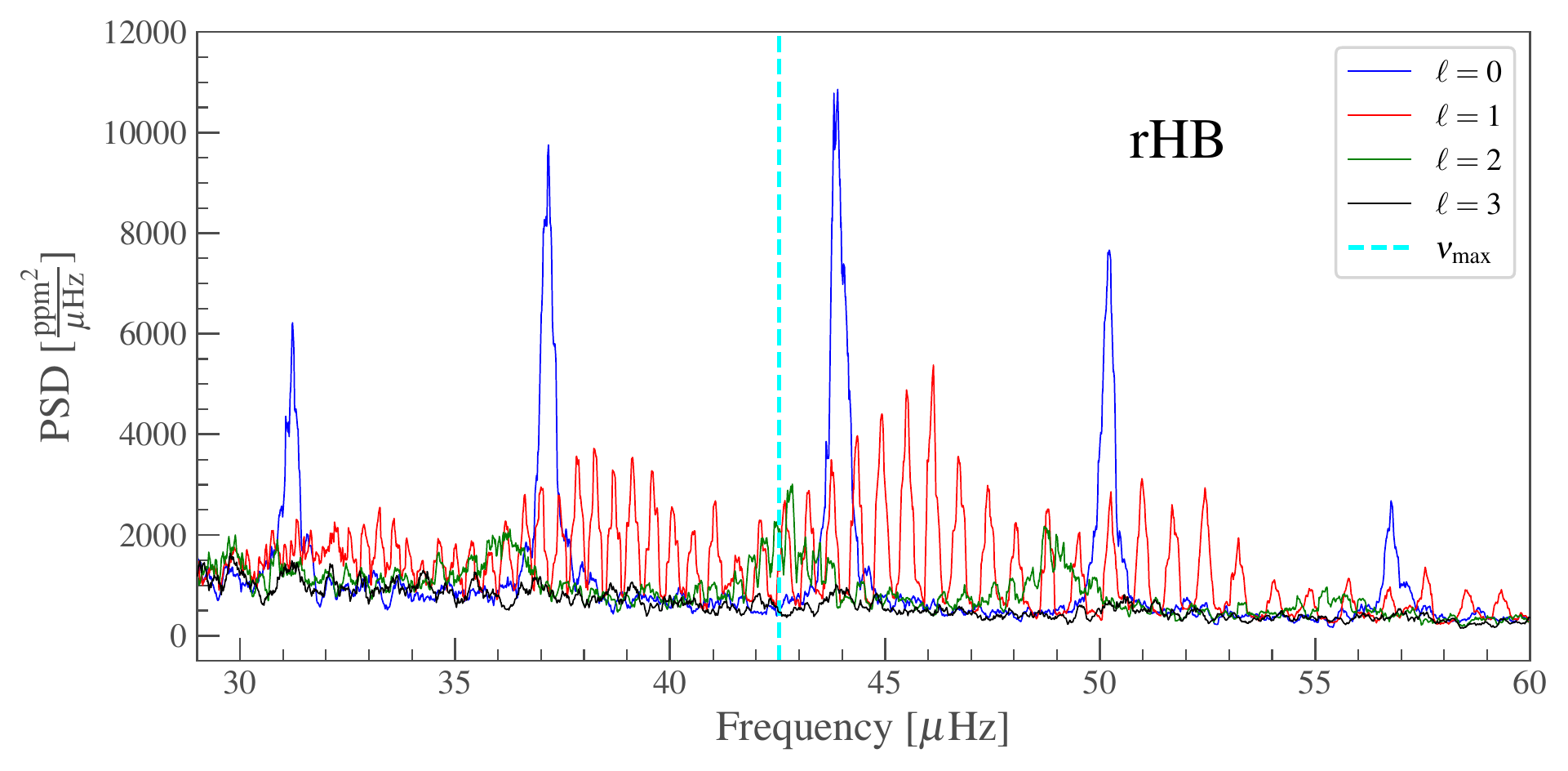}
\includegraphics[width=\columnwidth]{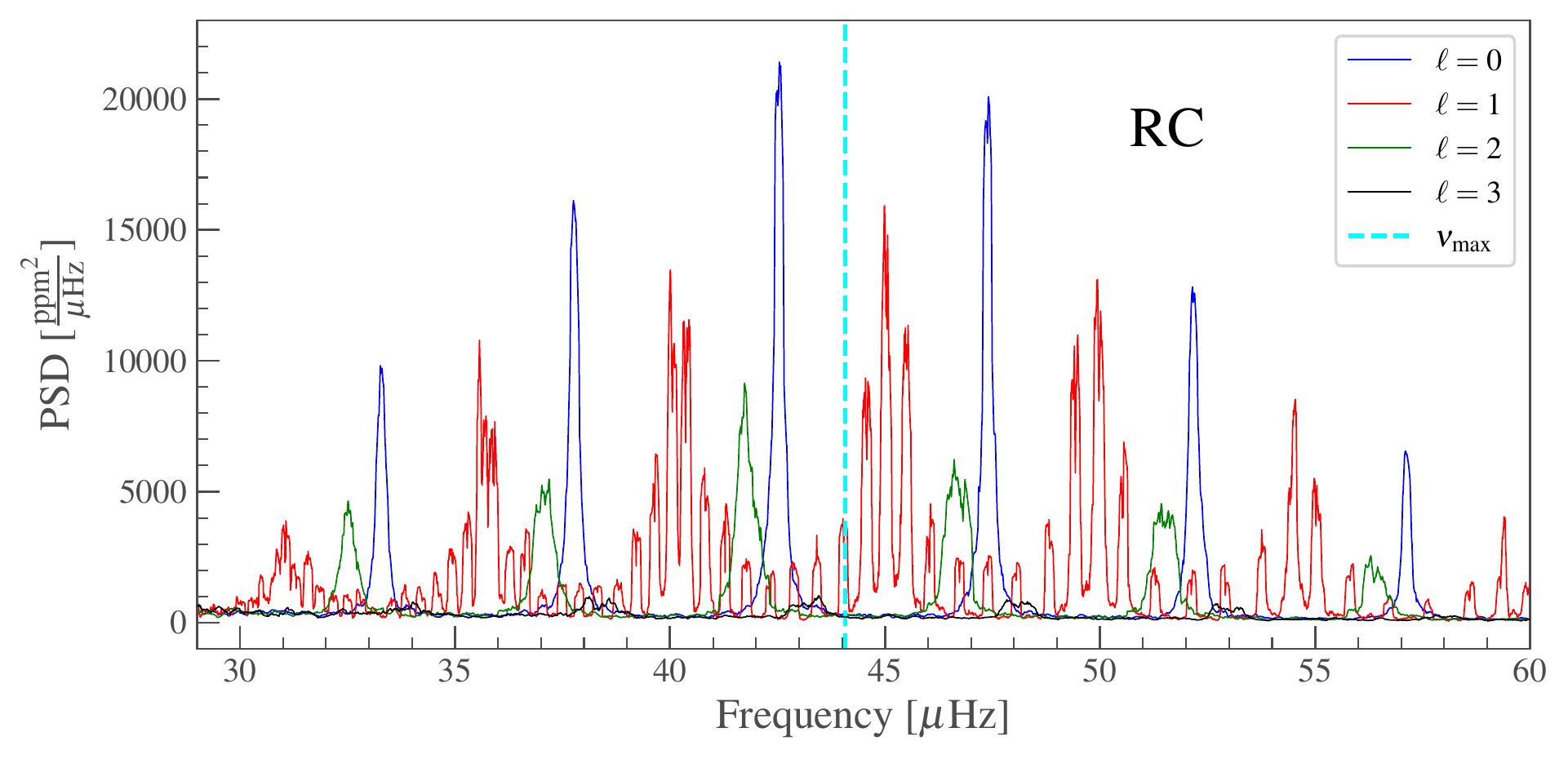}
%%%%%%%%%%%%%%%%%%%%%%%%%%%%%%%%%%%%%%%%%%%%%%%%%%%%%%%%%%%%%%%%%%%%%%%%%%%%%%%%%%%%%%%%%%%%%%%%
\caption{Smoothed version of the PSDs presented in Section~\ref{sec:psd}. Here we show the individual degrees for the rHB (top) and RC (bottom) simulated stars. The dashed cyan line is the corresponding \numax.}
\label{fig:eigen_inertia_psd_smooth}
\end{figure}

In this section we break down the PSDs of our reference models (Fig.~\ref{fig:eigen_inertia_psd})  into the contributions from the modes of the different angular degree. The smoothed PSDs for $\ell=0, 1, 2, 3$ are shown in Fig.~\ref{fig:eigen_inertia_psd_smooth}. The smoothing is chosen just for showing purposes, i.e. to resemble a Lorentzian fit of each eigenmode. The modulation around the p-like mode in the dipole modes of the RC star and the higher number of observed mixed modes in the rHB model are evident. Furthermore, the quadrupole modes of the rHB model are less visible than those of the RC model, and its octupole modes resemble a continuous background with small peaks almost coinciding with the radial modes.
Finally, we want to notice that the presence, in rHB stars, of $\ell =1,2,3$ modes very close to the radial ones (in some cases almost coinciding, e.g. Fig.~\ref{fig:eigen_inertia_psd_smooth}) could introduce a non-negligible influence on the analysis of the heights and the linewidths of the $\ell = 0$ modes.
%This large number of dipole modes in rHB stars could introduce a non-negligible influence on the analysis of the heights of the radial modes as they are also present close to them.
%This non-negligible influence, in rHB stars, of the mixed modes located very close to the position of the radial modes position could lead to a bias in the analysis of the background, and in the analysis of the heights of the radial modes.

\end{appendix}

\end{document}